\documentclass[fleqn,usenatbib]{mnras}

\usepackage{newtxtext,newtxmath}

\usepackage[T1]{fontenc}

\DeclareRobustCommand{\VAN}[3]{#2}
\let\VANthebibliography\thebibliography
\def\thebibliography{\DeclareRobustCommand{\VAN}[3]{##3}\VANthebibliography}

\usepackage{graphicx}	
\usepackage{relsize}    
\usepackage{float}

\title[Squicciarini et al. : The Scorpion's tale]{Unveiling the star formation history of the Upper Scorpius association through its kinematics}

\author[Squicciarini et al.]{
Vito Squicciarini$^{1,2}$,\thanks{E-mail: vito.squicciarini@inaf.it}
Raffaele Gratton$^{2}$, Mariangela Bonavita$^{2,3,4}$, Dino Mesa$^{2}$
\\
$^{1}$Department of Physics and Astronomy "Galileo Galilei", University of Padova, Italy \\
$^{2}$INAF -- Osservatorio Astronomico di Padova, Vicolo dell'Osservatorio 5, I-35142 Padova, Italy \\
$^{3}$ Scottish Universities Physics Alliance (SUPA), Institute for Astronomy, University of Edinburgh, Blackford Hill, Edinburgh EH9 3HJ, UK \\
$^{4}$Institute for Astronomy, University of Edinburgh, EH9 3HJ, Edinburgh, UK
}

\date{Accepted XXX. Received YYY; in original form ZZZ}

\pubyear{2021}

\begin{document}
\label{firstpage}
\pagerange{\pageref{firstpage}--\pageref{lastpage}}
\maketitle

\begin{abstract}
Stellar associations can be discerned as overdensities of sources not only in the physical space but also in the velocity space. The common motion of their members, gradually eroded by the galactic tidal field, is partially reminiscent of the initial kinematic structure. Using recent data from Gaia EDR3, combined with radial velocities from GALAH and APOGEE, we traced back the present positions of stars belonging to Upper Scorpius, a subgroup of Scorpius-Centaurus, the nearest OB association. About one half of the subgroup (the "clustered" population) appears composed of many smaller entities, which were in a more compact configuration in the past. The presence of a kinematic duality is reflected into an age spread between this younger clustered population and an older diffuse population, in turn confirmed by a different fraction $f_D$ of disc-bearing stars ($f_D = 0.24\pm0.02$ vs $f_D = 0.10\pm 0.01$). Star formation in Upper Scorpius appears to have lasted more than 10 Myr and proceeded in small groups that, after a few Myr, dissolve in the field of the older population but retain for some time memory of their initial structure. The difference of ages inferred through isochrones and kinematics, in this regard, could provide a powerful tool to quantify the timescale of gas removal.
\end{abstract}




\section{Introduction}

The discipline of star formation, in all of its aspects, has always been of paramount importance in astrophysics: lying at the interception of several fields of studies, it stretches from the grandest scale of galactic evolution, passing through the intricate web of the interstellar medium, to the dusty scenery of primordial discs, where stellar furnaces mildly begin to shimmer and glow. The presence of active stellar formation in our Galaxy \citep{1999ARAA..37..311E} provides the opportunity to gaze at the process while it unfolds before our eyes: during the last years, several surveys have been performed to study the environment of  galactic star-forming regions \citep[e.g.,][]{2009PASP..121..213C}, the formation of multiple systems \citep[e.g.,][]{2011ApJ...731....8K}, or even to directly test the predictions from models of planetary formation \citep[e.g.,][]{2021aa...646A.164J}.

It has long been known \citep{1954LIACo...5..293A} that, following the collapse and fragmentation of gigantic structures called {\it{molecular clouds}}, a plethora of stars ($N=10$ to $10^5$) begins to form; initially concealed by the same dusty envelope which they are born from, they rapidly \citep[2-7 Myr;][]{2021MNRAS.504..487K} divest themselves of it by means of harsh stellar winds and ionising radiation, mostly originating from massive stars; HII regions, the impressive product of the irremediable alteration of the original cloud, are ruthlessly sculpted by the injection of energy and momentum from exploding supernovae \citep{2020MNRAS.498.4906B}; after just a few Myr, the region is virtually devoid of its original gas reservoir \citep{2001MNRAS.321..699K}. The abrupt change within stellar natal environment is proven by observations showing that, while at $t<5$ Myr stars are often still embedded in their parent cloud, after 10 Myr only $\sim 10\%$ of stars are found in bound clusters \citep{2003ARAA..41...57L}. What we witness as an {\it{association}} for just --in cosmic terms--  the blink of an eye \citep[10-100 Myr,][]{2016EAS....80...73M}, is therefore a young system, still dwelt by bright ephemeral OB stars and regions of active stellar formation \citep{2001MNRAS.321..699K}.

The kinematic signature of members of clusters and associations, first recognized in the Hyades cluster and in Ursa Major already in the 19th century \citep{1869RSPS...18..169P}, is slowly eroded as the galactic differential rotation and tides spread the stars, turning them into moving groups or streams \citep{2002ASPC..285..442L}. Given that the observed densities of associations are too low to give rise to significant close encounters and scatterings, their initial velocity structure can largely be conserved over the timescale of several Myr \citep{2018MNRAS.476..381W}, as the above-mentioned perturbation induced by the galactic tidal field is expected to begin dominating on timescales of $\sim 10^7$ yr \citep{2018MNRAS.476..381W}. If this is the case, we might think to use our present knowledge of an association to delve into its past. The first attempt in this direction was done by \cite{1946PGro...52....1B,1964ARaa...2..213B}, who devised a simple linear expansion model, where all the sibling stars move away at a constant pace from their natal position. By tracing back their motion, it is in principle possible to obtain an estimate of the association age in a way that is independent of stellar evolution models. However, a quantitative assessment of the expansion model has long been considered elusive due to difficulties, on the one hand, in distinguishing real members from interlopers and, on the other hand, to obtain precise measurements of stellar distances and motions. This is why the idea has largely been shelved for decades \citep{1997MNRAS.285..479B}, even though the notion of OB associations as the inflated outcome of compact clusters kept being popular \citep{2003ARAA..41...57L}.

Ultra-precise astrometry from Gaia \citep{2016aa...595A...1G}, covering, in its last release, almost 2 billion sources all over the sky, providing reliable optical photometry up to $G\approx 21$ mag and, most importantly, supplying the astronomical community with astrometry and proper motions of unprecedented accuracy, has been revolutionising our knowledge of the Galaxy. An extensive search for new members of the known associations has been undertaken, reaching, for the nearest ones, the hydrogen-burning limit \citep{2018ApJ...862..138G}, and opening up a kaleidoscope of opportunities for kinematic studies. Especially when combined with radial velocities from external catalogues, Gaia is able to delve deeper than ever into the core of association architectures, unearthing exquisite fragments of their history. Some examples include the Gamma Velorum cluster, showing two distinct kinematic components \citep{2014aa...563A..94J}, Taurus \citep{2017ApJ...838..150K}, Cygnus OB2 and its complex substructures \citep{2016MNRAS.460.2593W}, and even smaller structures like the TWA moving group \citep{2014aa...563A.121D}. OB associations, in particular, are spectacularly confirming the expectation that their kinematic substructure is reminiscent of its initial structure \citep[e.g.,][]{1981MNRAS.194..809L,2016MNRAS.460.2593W}. The same complexity emerges when studying the geometry and the internal motion of molecular clouds \citep{1991ApJ...378..186F,2013aa...554A..55H}: starting from structure analysis of prestellar cores \citep{2020aa...638A..74L} and from the variegated shapes taken by filaments and filamentary networks in the early phase of stellar formation \citep{2018aa...610A..77H,2021arXiv210404541H} that give rise to distinct stellar populations \citep[e.g.,][]{2019aa...621A..42A}, it is natural to think that the complex, fractal structure is inherited by young stars \citep{2001AJ....121.1507E,2008ApJ...674..336G}. 

In this regard, the nearest OB association to the Sun, Scorpius-Centaurus (Sco-Cen), naturally stands as an ideal benchmark. Spanning an enormous area of approximately $80^\circ \times 40^\circ$ in the sky, this very young \citep[$t < 20$ Myr,][]{2012ApJ...746..154P} region comprises a few regions still actively forming stars \citep{2008hsf2.book..351W}. The exquisite concoction of closeness, youth and low extinction, allowing detection of members down to the brown dwarf regime, has made it the target of many studies in the last decades \citep[e.g.,][]{1999AJ....117..354D,2002AAS...200.7114M} involving binaries \citep[e.g.,][]{2013ApJ...773..170J}, primordial discs \citep[e.g.,][]{2006ApJ...651L..49C}, high mass \citep{2012ApJ...756..133C}, Sun-like \citep[e.g.,][]{2012ApJ...746..154P} and low-mass stars \citep[e.g.,][]{2013MNRAS.431.3222L}, and even young planets \citep[e.g.,][]{2021aa...646A.164J}.

The star formation history of Sco-Cen is closely related to its spatial structure. Sco-Cen is classically divided into three main subgroups \citep{1946PGro...52....1B,1999AJ....117..354D}: going toward lower galactic longitude, Upper Scorpius (USCO), Upper Centaurus-Lupus (UCL) and Lower Centaurus-Crux (LCC). Both density and age have a spatial gradient, with USCO being more compact and younger than UCL and LCC \citep{2016MNRAS.461..794P}. This intriguing observation led  \cite{1999AJ....117.2381P} to put forward the idea of a triggered star formation, where the process, started in LCC, gradually expanded eastward by means of supernova shocks, causing star formation in USCO after some Myr. The same USCO is thought to have triggered a minor burst of star formation in the $\rho$ Ophiuchi complex, which is still ongoing \citep{2008hsf2.book..351W}. However, the picture is complicate, and the three subgroups appear composed of many smaller entities, each bearing a peculiar mark while being conditioned by feedback from the surrounding environment  \citep{2018MNRAS.476..381W}. This tension between nature and nurture has given renovated impulse to the idea of investigating the substructure of the three subgroups to gain knowledge into their star formation histories.

The presence of at least a certain degree of substructure is evident even to visual inspection in the youngest part of Sco-Cen, USCO. This rather compact \citep[98×24×18 pc$^3$,][]{2018MNRAS.477L..50G} region of Sco-Cen, home to the bright Antares \citep{2013aa...555A..24O}, received a great attention on its own due to the interplay of kinematic and age peculiarities. Notably, a consistent age determination for USCO has long been elusive. While the first photometric studies argued for an age of $\sim 5$ Myr with no significant spatial and temporal spread \citep{1999AJ....117.2381P,2002AJ....124..404P}, recent work has been increasingly prone to an older age ($t \sim 11$ Myr) with a significant spread \citep[$\Delta t \sim 7$  Myr,][]{2012ApJ...746..154P}; the debate on the dependence of the latter on position \citep{2016MNRAS.461..794P}, spectral class \citep{2016ApJ...817..164R}, systematic artefacts due to stellar models \citep{2016aa...593A..99F} or an extended star formation history \citep{2017ApJ...842..123F} has been vivid in recent years.

In this context, insight from kinematic studies are pivotal to shed light on the problem. While the first studies, limited to few bright members, could not but aim at assessing a single common expansion age \citep{1978ppeu.book..101B} --a solid lower limit of $\sim 10$ Myr, in this regard, was put by \cite{2012ApJ...746..154P}--, nowadays we do have the means to investigate the whole kinematic substructure of USCO. 

The paper is organized as follows: after defining the selection criteria for our sample of USCO stars, together with the astrometric, kinematic and photometric data employed throughout this work (Sect. ~\ref{section:data}), we introduce in Sect. ~\ref{section:analysis} our tool, \textsc{madys}, and apply it to the region to recover and characterize the dual kinematic substructure found within the association. Sect.~\ref{section:age_determination} is dedicated to the age determination of clustered and diffuse populations, conducted in a threefold way. In Sect. ~\ref{section:discussion}, we discuss our results within the framework of previous studies of the region, with particular emphasis on its star formation history. Finally, in Sect. ~\ref{section:conclusions} we provide a brief summary of the results of this work. Appendix ~\ref{section:excess_factor} and ~\ref{section:proj_effects} explore in greater detail two quantities introduced to correctly handle data coming from Gaia: namely, a quality cut defined to exclude unreliable $G_{BP}$ and $G_{RP}$ photometric measurements and a set of corrections to remove the fraction of individual proper motions due to the reflection of the relative motion of USCO with respect to the Sun.

\section{Data}
\label{section:data}

\subsection{Sample selection}
Motivated by the idea of exploiting the full potential of the latest Gaia release \citep[EDR3,][]{2020arXiv201201533G}, we decided to construct a novel sample of USCO sources, independent from the DR2-based samples already present in the literature \citep[e.g.,][]{2020AJ....160...44L}. A preliminary deep query was done in a region virtually encompassing the whole Upper Scorpius, employing just minimal cuts on astrometry ($\alpha$, $\delta$, $\pi$) and kinematics ($\mu_\alpha^*=\mu_\alpha \cdot \text{cos}(\delta)$, $\mu_\delta$) to exclude stars either from the field or belonging to the nearby Upper Centaurus-Lupus (UCL) subgroup (Tab.~\ref{tab:criteria}.). No attempt has been done to remove, as many previous studies, sources belonging to the nearby Rho Ophiuchi region (from this moment on, $\rho$ Ophiuchi) since we intend to explore in detail its relation with the bulk of USCO. We will simply refer to our sample as {\it{USCO}}.

Membership to USCO has been defined operationally, by inspecting the 5D phase space $(\alpha, \delta, \pi, v_\alpha, v_\delta)$, with

\begin{align}
v_\alpha \text{ [km s$^{-1}$]} &= A \cdot \mu_\alpha^*/\pi \\
v_\delta \text{ [km s$^{-1}$]} &= A \cdot \mu_\delta/\pi
\end{align}

\noindent
with $A=4.74$ km yr s$^{-1}$ being the conversion factor between AU yr$^{-1}$ and km yr s$^{-1}$. The line of sight velocities $v_\alpha$ and $v_\delta$ are more suitable than proper motion components in a region of non-negligible radial depth ($ \Delta r \sim 50$ pc) and with parallax uncertainties no more as limiting as it was in the pre-Gaia era.

\begin{figure*}
\centering
\includegraphics[width=17cm]{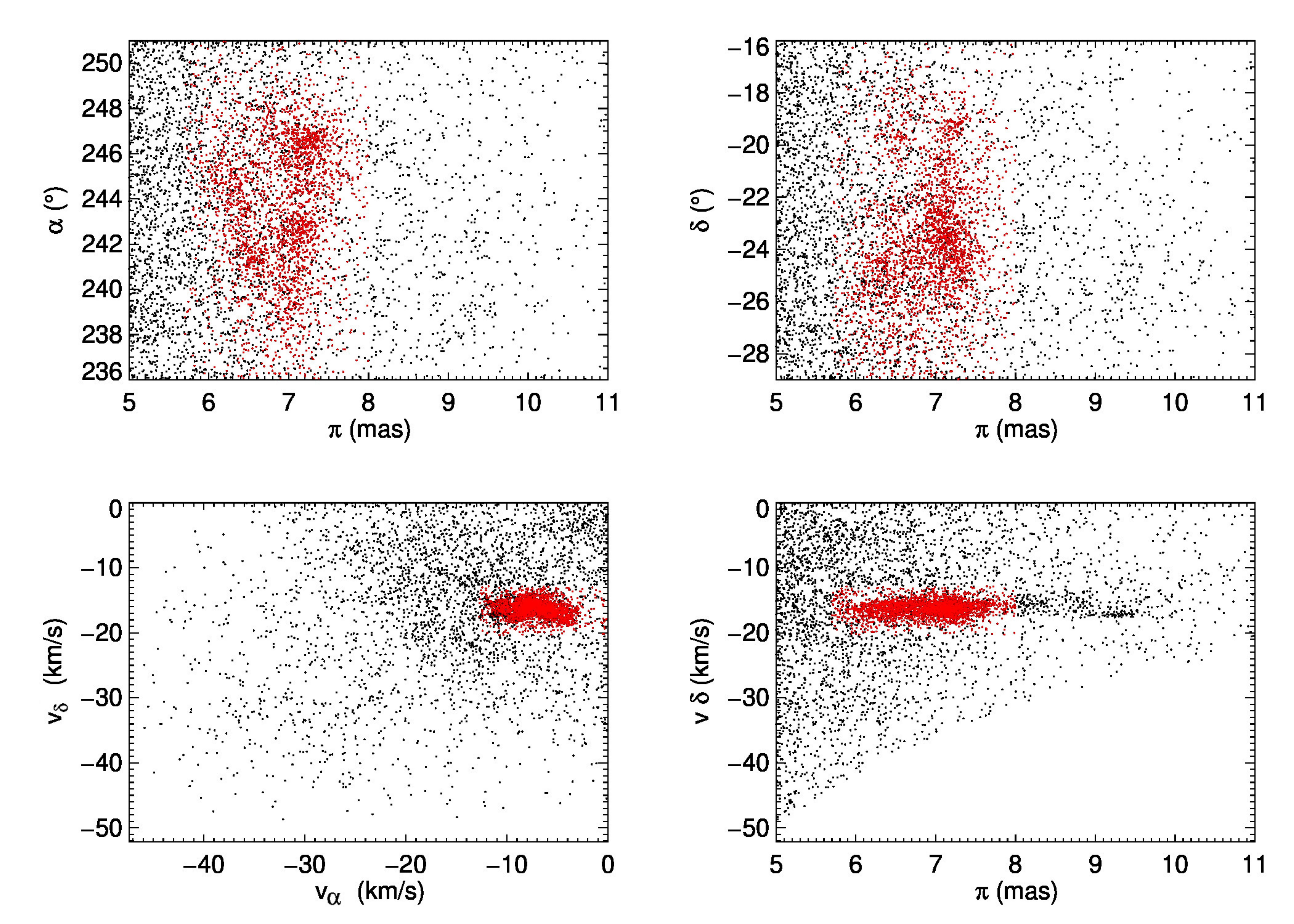}
\caption{Detection of USCO (red) within the 5D phase space. Only field stars (black) with $236^\circ<\alpha<251^\circ$, $-29^\circ<\delta<-16^\circ$, $G<20$ mag and $\sigma_\pi/\pi<0.1$ are shown for the sake of clarity.}
\label{fig:phase_space}
\end{figure*}

A clear concentration of sources emerges, distinguishing USCO from the field (Fig.~\ref{fig:phase_space}). A comparison with an independent sample, the DR2-based catalogue of Sco-Cen members by \cite{2019aa...623A.112D}\footnote{Actually, with the subsample defined by the same cuts on $\alpha$ and $\delta$ as our sample in order to exclude UCL members. Only their {\it{bona fide}} members were considered.}, yielded excellent agreement: out of their 2330 stars, 2129 ($\sim 91 \%$) were recovered; the fraction would have risen to 2298/2330 ($\sim 98\%$), if we employed their same cut on minimum distance ($\pi<10$ mas). However, we opted for a more conservative $\pi<8$ mas not to detrimentally affect field contamination.

\begin{figure*}
\centering
\includegraphics[width=17cm]{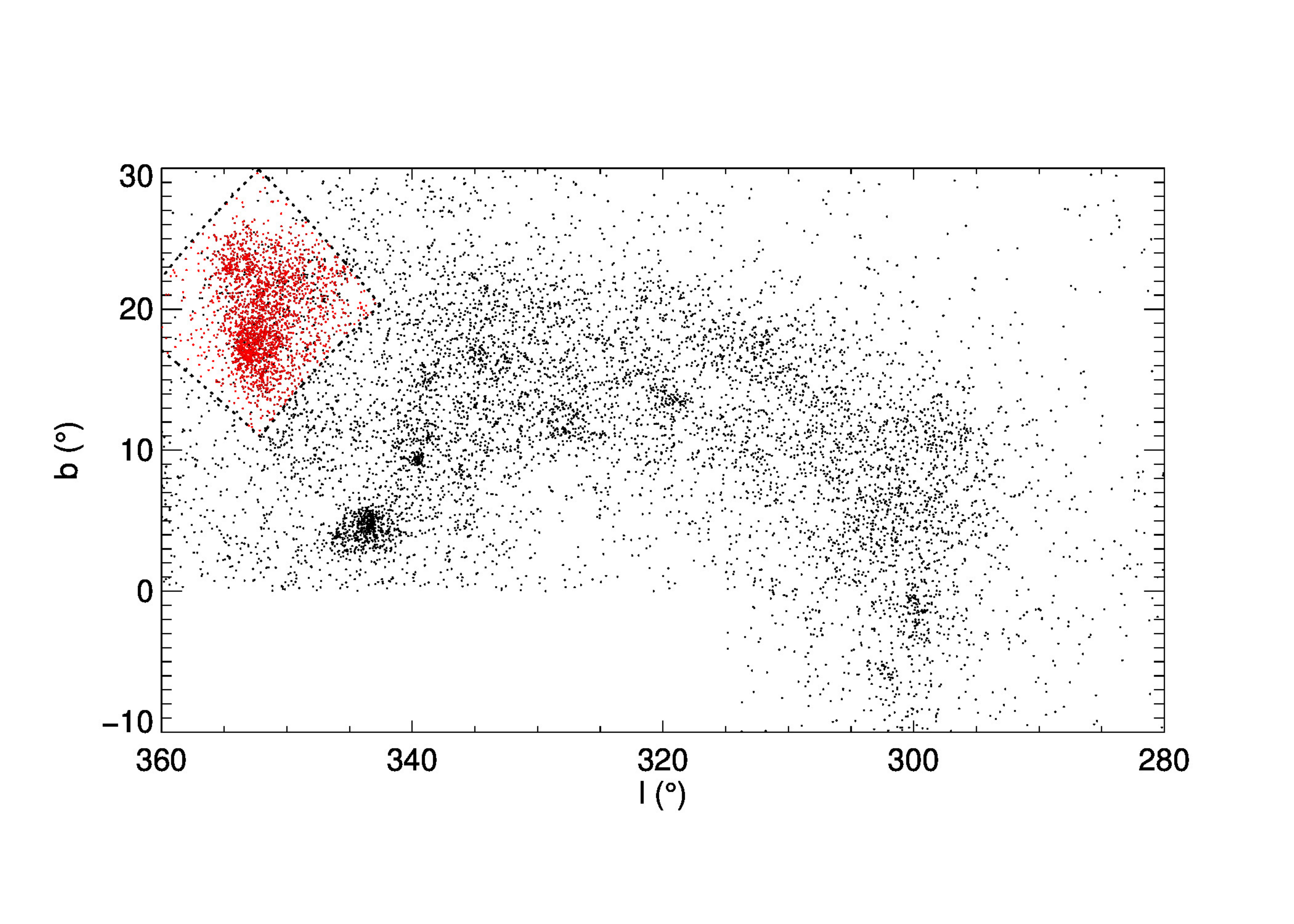}
\caption{Sco-Cen bona fide members from \protect\cite{2019aa...623A.112D}, shown in black. Upper Scorpius can easily be distinguished in the upper left. The sample used throughout this work and defined by the cuts of Table \ref{tab:criteria} is displayed in red. The criteria on right ascension and declination define the region bordered by the dashed lines.}
\label{fig:coord}
\end{figure*}

We consider those stars --the only additional caveats being $G<20$ and $\sigma_\pi / \pi < 0.1 $-- as our final sample (which we will call {\it{2D sample}}). The complete set of defining criteria are summarized in Table~\ref{tab:criteria}, while the sky distribution of the sample, comprising 2745 stars, is shown in Fig.~\ref{fig:coord}.

\begin{table}
  \centering
   \caption{Criteria for the selection of the 2D sample. Coordinates and proper motions are referred --as usual for Gaia EDR3-- to the ICRS at epoch J2016.0.}
   \begin{tabular}{lc} \hline \hline
\multicolumn{2}{c}{Initial query} \\
            \hline
  Query position $(\alpha_0,\delta_0)$ & $(245^\circ,-25^\circ)$ \\
  Query radius $(^\circ)$& 20 \\ 
  Parallax (mas) & $5<\pi<11$ \\
  Proper motion along $\alpha$ (mas yr$^{-1}$) & $-50<\mu_\alpha^*<0$ \\ 
  Proper motion along $\delta$ (mas yr$^{-1}$) & $-53<\mu_\delta^*<0$ \\ 
  No. of sources & 408465 \\ 
            \hline \hline
\multicolumn{2}{c}{Final criteria} \\
            \hline
  Right ascension $(^\circ)$ & $236<\alpha<251$ \\
  Declination $(^\circ)$ & $-29<\delta<-16$ \\
  Parallax (mas) & $5.7<\pi<8$ \\
  Parallax error & $\sigma_\pi/\pi <0.1$ \\
  Velocity along $\alpha$ (km s$^{-1}$) & $v_\alpha>-12.8$ \\ 
  Velocity along $\delta$ (km s$^{-1}$) & $-20.4<v_\delta<-12.8$ \\
  Apparent {\it{G}} magnitude (mag) & $G<20$ \\ 
  No. of sources & 2745 \\ 
  \hline\hline    
   \end{tabular}
   \label{tab:criteria}
\end{table}

\subsection{Radial velocities}
\label{section:RV}
An unbiased analysis of an extended region on the sky cannot be achieved, due to projection effects, without a full knowledge of the 6D phase space of its members: radial velocities (RV) are crucial not only to identify interlopers but also, more importantly, to correctly analyse stellar motions (see Appendix ~\ref{section:proj_effects} for details).

A complete analysis in the 6D phase space has been performed on the subsample possessing reliable RV measurements. In addition to Gaia EDR3, we collected data from APOGEE DR16 \citep{2020ApJS..249....3A} and GALAH DR3 \citep{2020arXiv201102505B}; whenever multiple measurements were present, the datum with the smallest error bar was chosen. After selecting sources with a relative error on RV < 0.1 or an absolute error $<1$~km~s$^{-1}$, we defined the Cartesian frame $(x,y,z)$:
\begin{equation} \label{eq:coord_transf1}
    \begin{cases}
    x=r \text{ cos} (\delta-\delta_P) \text{ sin} (\alpha-\alpha_P) \\
    y=r \text{ cos} (\delta-\delta_P) \text{ cos} (\alpha-\alpha_P) \\
    z=r \text{ sin} (\delta-\delta_P)
\end{cases}
\end{equation}
where the pole $(\alpha_P,\delta_P)=(243.09^\circ, -23.03^\circ)$ points, for convenience, toward the mean equatorial coordinates of the sample. Finally, we restricted only to sources with propagated errors on $v_x$, $v_y$ and $v_z$ simultaneously satisfying the three conditions\footnote{$1 \text{ km s}^{-1}=1.02 \text{ pc Myr}^{-1}$.}:
\begin{itemize}
    \item $|\sigma_{v_x}/v_x|<0.1\quad \text{OR}\quad \sigma_{v_x}<0.1 \text{ pc Myr}^{-1}$,
    \item $|\sigma_{v_y}/v_y|<0.1\quad \text{OR}\quad \sigma_{v_y}<0.1 \text{ pc Myr}^{-1}$,
    \item $|\sigma_{v_z}/v_z|<0.1\quad \text{OR}\quad \sigma_{v_z}<0.1 \text{ pc Myr}^{-1}$.
\end{itemize}
The final RV sample ({\it{3D sample}}) comprises 771 stars, $\sim 28\%$ of the 2D sample (Table~\ref{tab:data_errors}).

Although we decided not to appoint the 3D sample as our main focus, because it only imperfectly reproduces the real distribution of sources\footnote{The distribution of sources possessing RV does not appear as a random pick of the Gaia sample, but rather --as expected-- as the union of distinct surveyed regions.}, we will employ it in a twofold way: on the one hand, it will provide us with a way to quantify the effect of the association's mean motion with respect to the Sun (which we will call, from this moment on, {\it{bulk motion}}); on the other hand, it will offer a constant comparison with the 2D sample to check the validity of our results:
by comparing, whenever possible, features observed in 2D with their 3D counterparts, we are able to rule out the possibility of a random alignment of sources lying at different distances, i.e. a perspective effect.

As regards the former aspect, a simple geometrical argument proves that, taken a group of stars, the knowledge of their proper motions alone is not sufficient to disentangle between a real expansion and a non-zero mean radial motion with respect to the line of sight through its centre, i.e. a {\it{virtual expansion}} \citep[see discussion in][]{1999AJ....117..354D}. Transverse motions, too, are split up into velocity components depending on $(\alpha,\delta,\pi)$. Thus, any non-zero bulk motion will manifest itself as a bias in the kinematic reconstruction.

Once having estimated that the centre of the 3D sample approximately lies at $(\alpha_c,\delta_c,r_c)=(244.55^\circ,-23.79^\circ,143.3$ pc$)$, we computed the
mean velocity components in a Cartesian frame centred on it. Expressed in the standard right-handed Cartesian Galactic frame, our 3D sample has median velocity components $(U, V, W) = (-4.788 \pm 0.019, -16.378 \pm 0.015, -6.849 \pm 0.016)$ km s$^{-1}$, with a precision gain of almost one order of magnitude relative to previous estimates \citep{2012ApJ...758...31L,2018MNRAS.477L..50G}. Finally, we determined the projections of this bulk motion on the proper motions of each star of the 2D sample, and subtracted them (see Appendix ~\ref{section:proj_effects} for details). We verified that, due to the angular extent of USCO, this bias does not significantly affect the shape of the substructures nor the timing of their maximum spatial concentration.

Even correcting for bulk motion, a similar (although smaller) projection effect keeps affecting individual stellar velocities, due to the rotation of the $(v_\alpha, v_\delta, v_r)$ plane with $(\alpha,\delta)$. Again, the angular extent of the association is not too big to hinder the approach altogether\footnote{\label{foot:cartesian}. Choosing a fixed Cartesian $(\hat{x},\hat{y},\hat{z})$ frame around a star defined by $(\alpha_0, \delta_0, \pi_0, v_{\alpha,0}, v_{\delta,0}, v_{r,0})$ such that $\hat{x} \parallel v_{\alpha,0} $, $\hat{y} \parallel v_{r,0}$ and $\hat{z} \parallel v_{\delta,0}$, the mixing between velocity components for a second star having $(\alpha_1=\alpha_0+10^\circ,\delta_1=\delta_0+10^\circ, \pi_1, v_{\alpha,1}, v_{\delta,1}, v_{r,1})$ is such that $v_x = 0.985 v_{\alpha,1} - 0.030 v_{\delta,1} + 0.171 v_{r,1}$, $v_y = -0.174 v_{\alpha,1} -0.171 v_{\delta,1} +0.970 v_{r,1}$, $v_z=0.985 v_{\delta,1} +0.174 v_{r,1}$.}.

\begin{table}
  \centering
   \caption{Median uncertainties on astrometry and kinematics for the 2D and 3D samples.}
   \begin{tabular}{lclc} \hline \hline
\multicolumn{2}{c}{2D sample} & \multicolumn{2}{c}{3D sample} \\
            \hline
  No. of sources & 2745 & No. of sources & 771 \\
  Median $\sigma_\pi$ (mas) & $0.06$ & Median $\sigma_{x}$ (pc) & $0.03$ \\
  Median $\sigma_{\mu_\alpha^*}$ (mas yr$^{-1}$) & $0.07$ & Median $\sigma_{y}$ (pc) & $0.82$\\
  Median $\sigma_{\mu_\delta}$ (mas yr$^{-1}$) & $0.05$ & Median $\sigma_{z}$ (pc) & $0.03$ \\
  Median $\sigma_{v_\alpha}$ (km s$^{-1}$) & $0.08$ & Median $\sigma_{v_x}$ (pc Myr$^{-1}$) & $0.05$ \\
  Median $\sigma_{v_\delta}$ (km s$^{-1}$) & $0.14$ & Median $\sigma_{v_y}$ (pc Myr$^{-1}$) & $0.04$ \\ 
  & & Median $\sigma_{v_z}$ (pc Myr$^{-1}$) & $0.10$ \\
            \hline \hline
   \end{tabular}
   \label{tab:data_errors}
\end{table}

\subsection{Photometry}
\label{section:photometry}

Whereas astrometric and kinematic data were gathered from Gaia EDR3, the photometry comes from Gaia DR2 \citep{2018aa...616A...1G}: the reason is that the set of isochrones that we used relies on Gaia DR2, and the filter response is not exactly the same between the two releases\footnote{We verified that, for our initial sample of 408465 sources, $G_{BP}$ magnitudes are on average 0.3 mag dimmer in EDR3.}. As $G_{BP}$ and $G_{RP}$ photometry can be severely contaminated by the background at faint magnitudes \citep{2021gdr3.reptE...5B}, a quality cut was needed to discriminate whether $G_{BP}$ and $G_{RP}$ magnitudes could be considered reliable or not. Following the line of reasoning by \citet{2020arXiv201201916R}, a colour-independent {\it{BP-RP excess factor}} $C^*$ was defined, starting from the available bp\_rp\_excess\_factor C:
\begin{equation} \label{eq:excess}
C^* - C =
  \begin{cases}
    a_0+a_1\Delta G+a_2 \Delta G^2 +a_4 G & \text{ } \Delta G < 0.5 \\
    a_0+a_1\Delta G+a_2 \Delta G^2+a_3 \Delta G^3+a_4 G & \text{ } 0.5 \leq \Delta G < 3.5 \\
    a_0+a_1\Delta G+a_4 G & \text{ } \Delta G \geq 3.5
  \end{cases}
\end{equation}
where $\Delta G=(G_{BP}-G_{RP})$; the distribution of $C^*$ peaks at about 0 for well-behaved sources at all magnitudes but, when considering subsamples of stars with similar brightness, it tends to widen out for fainter G; a varying standard deviation $\sigma_{C^*}(G)$ can be defined as:
\begin{equation} \label{eq:excess_sigma}
\sigma_{C^*}(G)=k_1+k_2 \times G^{k_3}
\end{equation}

Setting a rejection threshold at $3\sigma$, we labelled 889/2745 sources ($32\%$) as having unreliable ($G_{BP}$, $G_{RP}$) magnitudes; the effect is larger, as expected, at fainter ($G \gtrsim 15$) magnitudes. Details on the derivation of $C^*$ and $\sigma_{C^*}$, as well as the numerical values of the constants of Eq.~\ref{eq:excess}-~\ref{eq:excess_sigma} can be found in Appendix ~\ref{section:excess_factor}.

Infrared measurements from 2MASS \citep{2006AJ....131.1163S} were gathered and cross-matched with Gaia magnitudes, by inspecting the radial distance $r$ between nearby Gaia and 2MASS sources. About $96\%$ of the source pairs have $r<0.7''$, so this value was chosen to establish whether a pair actually referred to the {\it{same}} source. Only measurements labelled by the best quality flag ("A") were used.

In order to study the fraction of disc-bearing stars, additional photometric data were collected from WISE \citep{2010AJ....140.1868W} and ALLWISE \citep{2014yCat.2328....0C}. If simultaneously present, ALLWISE magnitudes were preferred over WISE data. As with 2MASS data, only measurements with the best quality flag ("0") were employed; measurements with $W1<8.1$ mag, $W2<6.7$ mag, $W3<3.8$ mag and $W4<-0.4$ mag were not considered, due to saturation problems at bright magnitudes \citep{2012wise.rept....1C}. The effect of flux contamination from nearby sources, increasingly affecting surveys with longer wavelengths, will be assessed in Sect. ~\ref{section:disc_fraction}.

\subsection{Extinction}

\begin{figure}
\centering
\includegraphics[width=9cm]{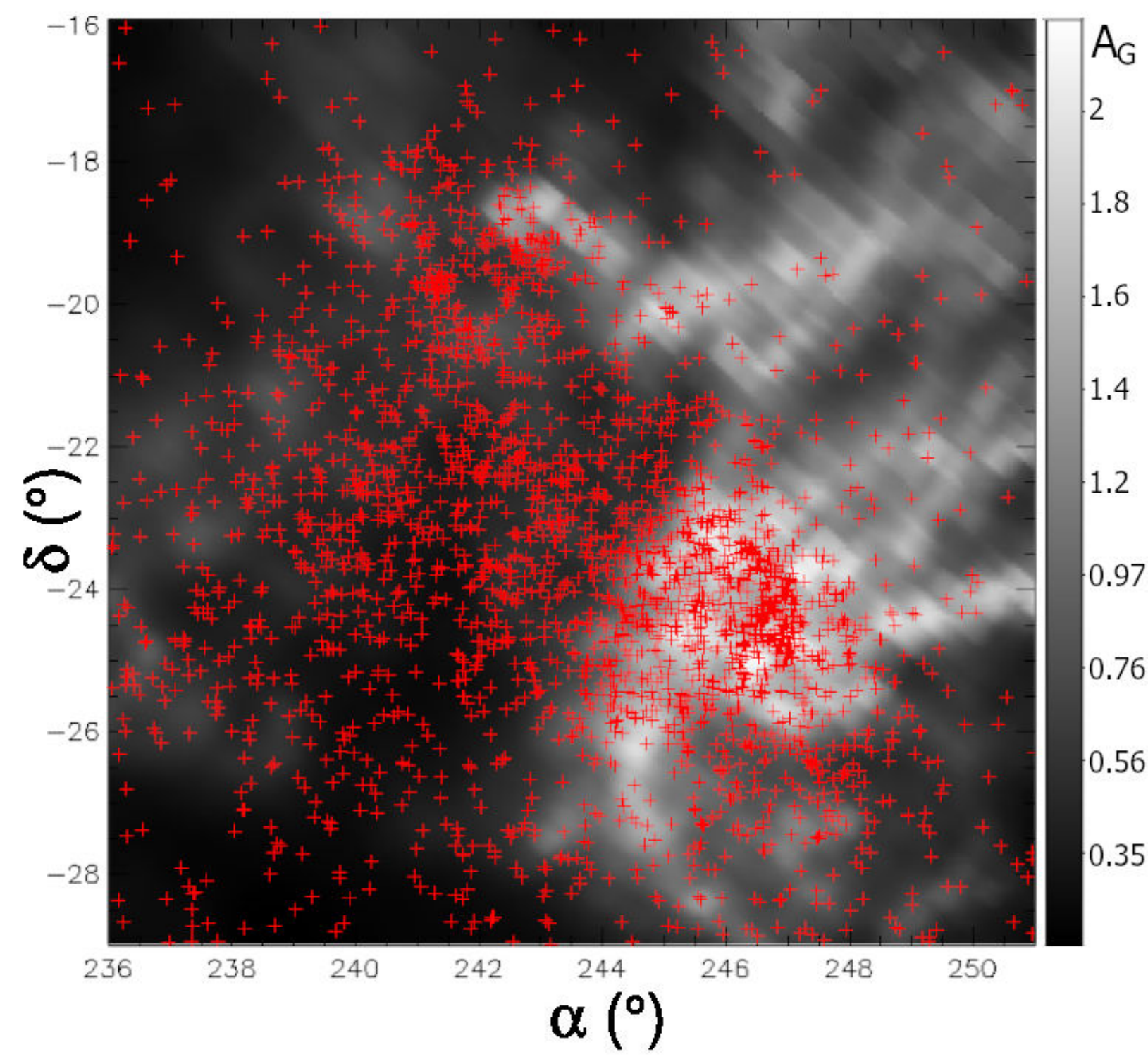}
\caption{Line of sight G-band extinction (mag) in the direction of USCO, plotted with the help of a colour bar. The 2D sample is overplotted in red. The values, derived from the map by \protect\cite{2020aa...639A.138L}, are computed for simplicity at a fixed $d=160$ pc, encompassing $\sim 90\%$ of the sample.}
\label{fig:extinction}
\end{figure}

Despite the proximity of the region, interstellar extinction is not to be overlooked: the youngest region here studied, the $\rho$ Ophiuchi cloud, can reach $A_V \sim 40-50$ mag in its core \citep{1983ApJ...274..698W}, preventing detection of its embedded protostars \citep{2021arXiv210112200G} in optical surveys, while the corners of the association hardly reach $A_V \approx 0.2$ mag. In order to take such intrinsic spatial variability into account, the absolute photometry of each star was corrected via the interstellar extinction map by \cite{2020aa...639A.138L}. The resolution of the map is 1 pc, which at $d \sim 140$ pc typically translates into an angular resolution $\Delta \theta = 0.4^\circ$, comparable to the extent of the largest molecular clouds within $\rho$ Ophiuchi \citep{2016aa...588A.123R}. The provided G-band extinction was converted, case by case, to the appropriate band using a total-to-selective absorption ratio $R=3.16$ and extinction coefficients $A_\lambda$ taken from \cite{2019ApJ...877..116W}. A sketch of the integrated G-band extinction at a constant distance of $d=160$ pc is shown in Fig.~\ref{fig:extinction}.

\section{Analysis}
\label{section:analysis}
We developed a flexible Python tool, \textsc{madys}, that, given a list of stars, automatically retrieves and cross-matches photometry from different catalogues, corrects for interstellar extinction, and employs photometric data judged reliable to get ages and masses of individual stars by comparison with pre-MS isochrones. Additionally, \textsc{madys} can look for evidence of primordial discs and perform a group kinematic analysis. The tool, to be shortly made publicly available, constitutes a development of the age determination tool already employed in \cite{2021aa...646A.164J}. We describe in this Section the results of the kinematic analysis of our samples.

\subsection{Kinematic subgroups}
\label{section:subgroups}

\begin{figure*}
\centering
\includegraphics[width=17cm]{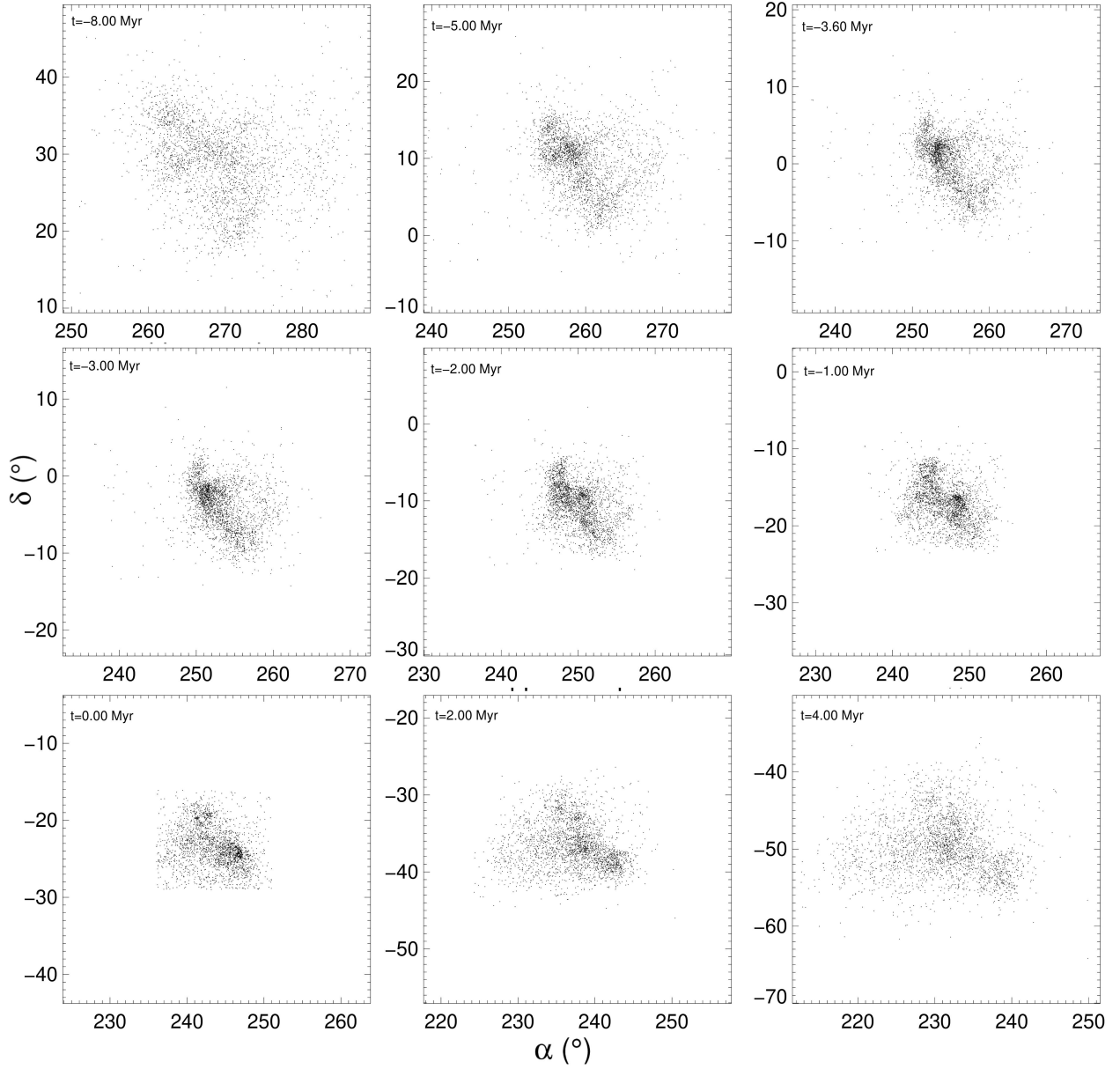}
\caption{A few frames of the time evolution of Upper Scorpius. Some clear overdensities of sources emerge at different times.}
\label{fig:time_frames}
\end{figure*}

As hinted in the previous sections, our method works by tracing back the celestial coordinates of the 2D sample on the basis of their present $\mu_\alpha$ and $\mu_\delta$, under the assumption that dynamical interactions among members are negligible. The inspection of the time evolution of the sample\footnote{Available in the supplementary material as a .mp4 movie.} in the $(\alpha,\delta)$ plane suggests that a wealth of information is still encapsulated beneath the present structure of USCO: the sight of different parts of the association, clustering at different times in the past, was what encouraged us to tentatively distinguish subgroups based on a purely kinematic way (Fig.~\ref{fig:time_frames}).

Our 2D sample was inspected via a semi-automated approach based on iterative k-means clustering \citep{2183c8d766ec4e20acbe35c017632c5c}. The analysis takes place in $n$ 4D planes $(\alpha(t),\delta(t), v_\alpha, v_\delta)$:
\begin{align}
\alpha(t) &= \alpha_0 + \frac{\mu_\alpha^*}{\xi \text{ cos}(\delta)} \\
\delta(t) &= \delta_0 + \frac{\mu_\delta}{\xi}
\end{align}

\noindent
where ($\alpha_0$, $\delta_0$) are the present coordinates, $\xi=3.6 \frac{\text{deg Myr$^{-1}$}}{\text{mas yr$^{-1}$}}$ is a factor needed to express angular velocities in units of [deg Myr$^{-1}$] and $n$ represents the number of time steps of an evenly-spaced temporal grid ($t \in [-15,15]$ Myr,  $\Delta t=0.2$ Myr). Every time a coherent group was visually identified at a certain $t$, it was subtracted from the sample. The procedure, which we will refer to as {\it{2D analysis}}, was iterated, until no additional groups could be found with confidence. 

\begin{figure}
\centering
\includegraphics[width=9cm]{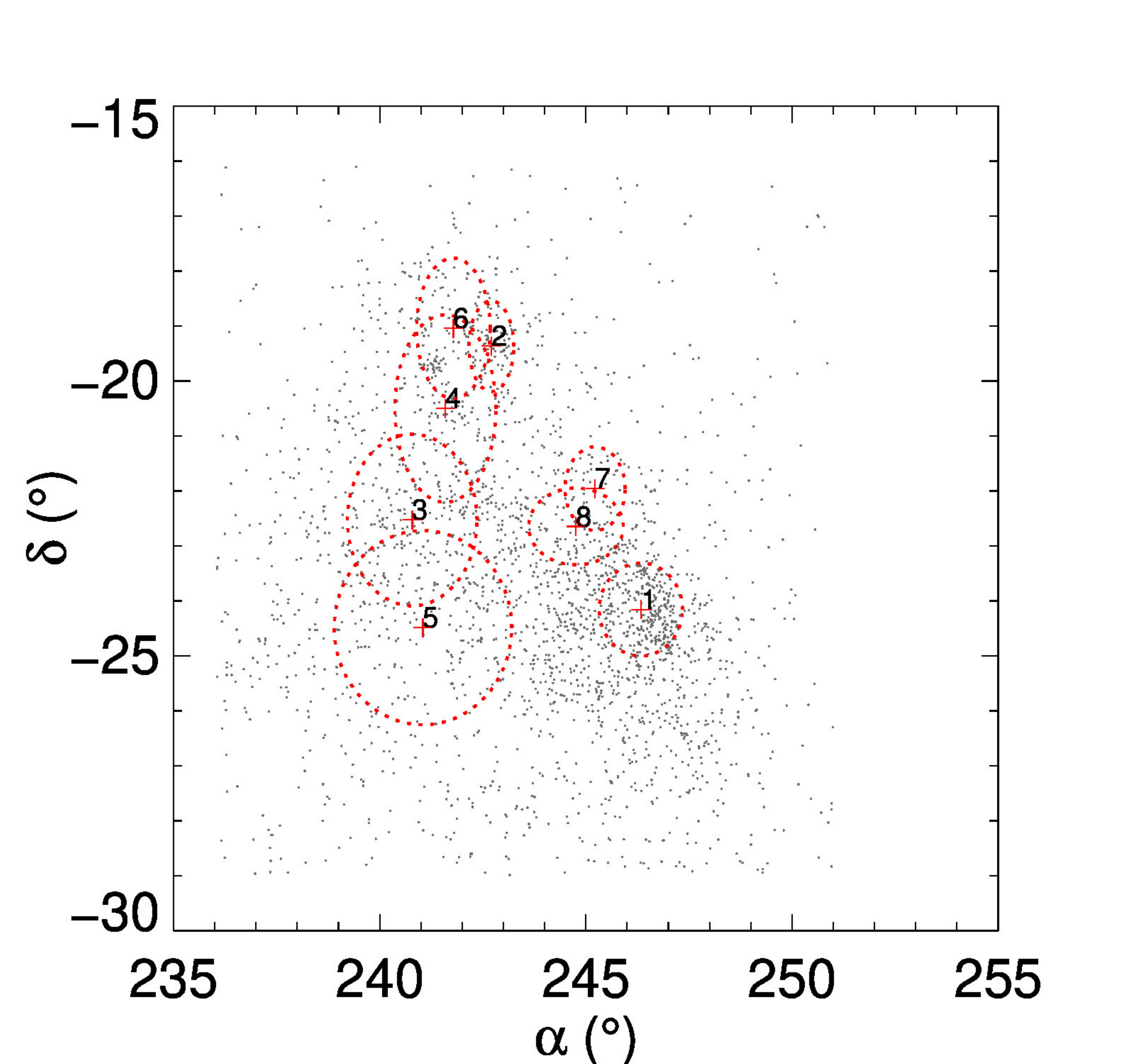}
\caption{Mean positions of the groups at present time, with $1 \sigma$ ellipses shown as dashed red curves.}
\label{fig:group_coord}
\end{figure}

\begin{figure*}
\centering
\includegraphics[width=17cm]{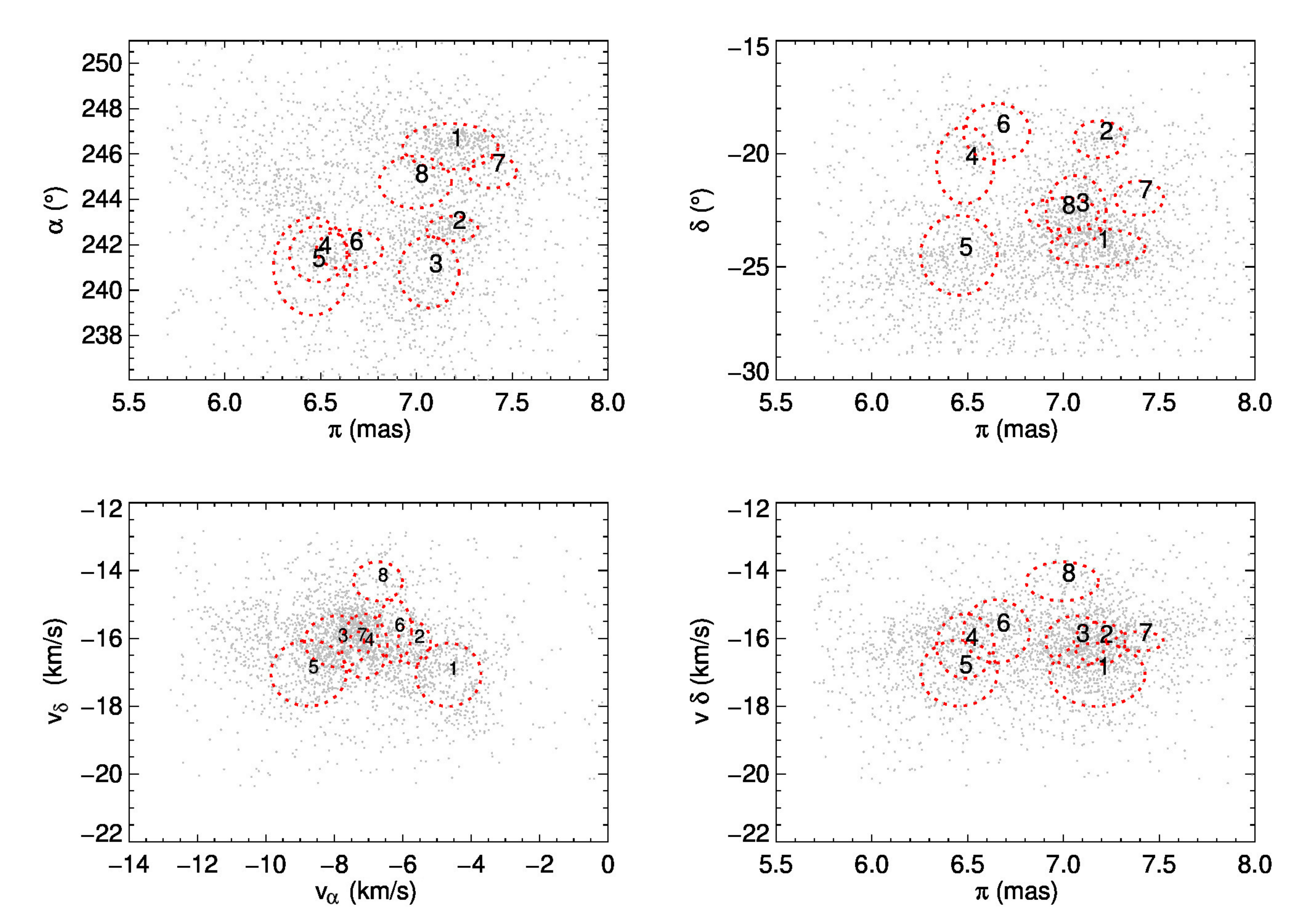}
\caption{Distribution of the groups in the 5D phase space, with $1 \sigma$ ellipses shown as dashed red curves.}
\label{fig:group_phase_space}
\end{figure*}

A total of 8 subgroups was identified, each with a peculiar mark in the phase space (Fig.~\ref{fig:group_coord}-~\ref{fig:group_phase_space}); their main properties are summarized in Table~\ref{tab:group_prop}, and include an estimate of the moment of maximum coherence $t_K$, which we will define in Sect.~\ref{section:kin_age}.

To get an estimate of the false alarm probability for these overdensities, i.e. to rule out that similar features could be produced by chance, we set up, for each subgroup, the following test: we built reshuffled 2D samples by randomly assigning to each star a quintuplet $(\alpha ,\delta , \pi, \mu_\alpha, \mu_\delta)$, every parameter being drawn independently from its natal distribution; then, we traced back the positions at the appropriate $t_K$, and excluded the most distant star --with respect to the 5D normal distribution-- in an iterative way, until as many stars as in the original subgroup were left. Then, we estimated the false alarm probability (2DP) as the ratio between the number of simulations ending up --at any time-- with a more clustered group than the observed one and the total number of performed simulations:
\begin{equation} \label{eq:real_prob}
\text{2DP} \approx f=\frac{N_{good}}{N_{good}+N_{bad}}
\end{equation}
We considered as a {\it{success}} a simulation yielding a median angular distance from its centre and a 1$\sigma$ velocity ellipse $\sigma_{v_\alpha}\cdot \sigma_{v_\delta}$ smaller than the observed ones. Similarly, a 3D false alarm probability (3DP) was computed by assigning to each star a sextuplet $(x,y,z,v_x,v_y,v_z)$ and considering as {\it{success}} a simulation yielding a smaller median distance from the centre than the real one and a 1$\sigma$ velocity ellipse\footnote{Computed, in the 3D case, from the samples restricted to the [16,84] percentiles to minimize the impact of possible outliers.} $\sigma_{v_x}\cdot \sigma_{v_y}\cdot \sigma_{v_z}$ smaller than the observed ones.

As already mentioned in ~\ref{section:RV}, the results is not influenced significantly by removal of projection effects due to USCO's bulk motion: the distinction among subgroups was carried out on the original 2D sample, using the corrected velocities as a tool to estimate their $t_K$. Also, errors on $v_\alpha$ and $v_\delta$, increasing linearly with time, cannot be the source of the pattern observed, as $95 \%$ of the sources has error $<0.5^\circ$ at $t=-5$ Myr.

We find strong evidence for the physical nature of groups 1, 2, 3, 4, 5 and the tiny group 7. While Group 1 corresponds to the $\rho$ Ophiuchi region, as confirmed by a cross-match with the catalogue by \cite{2019aa...626A..80C}\footnote{Out of 517 stars that are both in their sample and in our 2D sample, 342 (66\%) have been assigned to Group 1. Most of the remaining stars have not been assigned to any group.}, Group 2 and Group 3 visually resemble Group E and Group H of a recent work by \cite{2021arXiv210509338K}. The latter, an elongated structure spanning $\sim 1^\circ \times 3^\circ$ with a North-South orientation, is remarkable in many aspects: at a mean distance of $141\pm4$ pc, its shorter side, somewhat broadened by proper motion uncertainties (the $1\sigma$ angular error at $t=-3.5$ Myr is $\sim 0.15^\circ$), spans just $\sim 2 \text{ pc}$.

The randomness of groups 6 and 8 --for which the 3D sample is quite small-- cannot be ruled out. For the moment being, we will keep on retaining the division in subgroups, which will be further discussed in Sect. ~\ref{section:discussion}.

\subsection{Clustered and diffuse populations}

An even more interesting result is obtained if we put together the stars belonging to the subgroups, and compare the resulting population with the remaining stars. We might call them the {\it{clustered population}} (1442 stars) and the {\it{diffuse population}} (1303 stars), respectively. The difference between the two populations is profound: while the former appears to clump, both in 2D and in 3D, at different times in the past, the latter does not.

Since the clustered population has been created by assembling subgroups detected individually during the 2D analysis, the result is somewhat surprising. The 3D sample strengthens the idea of a kinematic duality within the association, that we tried to quantify by taking the median distance from the mean 3D position for the whole sample and for the two subpopulations:

\begin{equation} \label{eq:median}
 d_M(t)=\text{median} \left( \sqrt{[x(t)-\bar{x(t)}]^2+[y(t)-\bar{y(t)}]^2+[z(t)-\bar{z(t)}]^2} \right )
\end{equation}

\begin{figure}
\centering
\includegraphics[width=9cm]{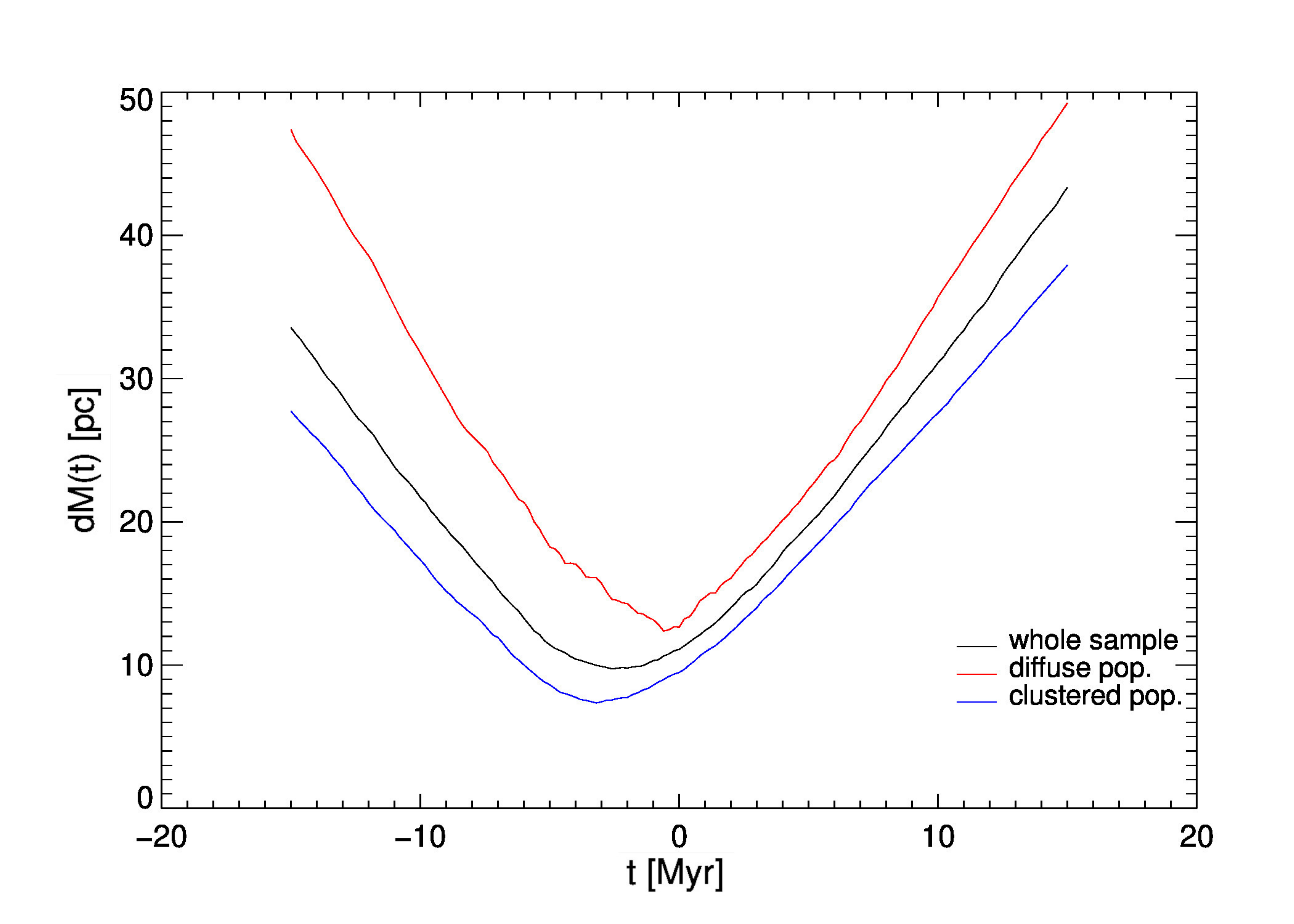}
\caption{The trend of $d_M(t)$, defined by Eq.~\ref{eq:median}, for the clustered and the diffuse population. The clustered population reaches a minimum $d_M \approx 7.4$ pc at $t=-3.2$ Myr, while the diffuse population does not appear not to have been more clustered in the past.}
\label{fig:median_clustered}
\end{figure}

Looking at Fig.~\ref{fig:median_clustered}, it is clear that $\tilde{d}_M := \text{min} (d_M(t))$ is smaller for the clustered population ($\tilde{d}_M = 7.4\pm0.1$ pc) than for the diffuse population ($\tilde{d}_M= 12.4\pm 0.2$ pc), and reaches its minimum value earlier for the former ($t=-3.2$ Myr) than for the latter ($t=0$ Myr). Errors are computed via a Monte Carlo simulation for uncertainty propagation, i.e. by repeating the procedure while randomly varying, in a normal fashion, input velocities according to their uncertainties.

While the moment minimising $d_M(t)$ cannot be assumed as an age estimate but rather as a lower limit (see Sect.~\ref{section:discussion}), its minimum value is indeed a measure of the degree of concentration of the population at that time. To shed light on the significance of its difference between the clustered and diffuse population, we ran the same set of 3D simulations described in Sect.~\ref{section:subgroups}, using $t_K=3.2$ Myr. Out of 100000 simulations, none behaved better than the original sample; the best-fitting Gaussian distribution of their $\tilde{d}_M(t)$, defined by $(\mu,\sigma)=(9.68 \text{ pc}, 0.16 \text{ pc})$, places a confidence level on the observed clustering at $\sim 14 \sigma$.

\begin{table*}
  \centering
   \caption{Number of stars (n) and mean positions in the phase space of the groups. Errors should be read as sample standard deviations, equivalent to the semimajor axes of the ellipses in Fig.~\ref{fig:group_phase_space}. The brightest star of each group (MRS) is indicated too.}
   \begin{tabular}{lcccccccccc} \hline \hline
Group & 2D n & 3D n & 2DP & 3DP & $\alpha$ & $\delta$ & $\pi$ & $v_\alpha $ & $v_\delta$ & MRS \\
 & & & & & deg & deg & mas & km s$^{-1}$ & km s$^{-1}$ & \\
            \hline

1 & 467 & 121 & $<0.1 \%$ & $<0.1 \%$ & $ 246.3\pm  1.0$ & $  -24.2\pm 0.8$ & $  7.2\pm 0.2$ & $ -4.7\pm 1.0$ & $ -17.1\pm 0.9$ & i Sco \\
2 & 114 & 64 & $<0.1\%$ & $<0.1\%$ & $ 242.7\pm 0.5$ & $ -19.4\pm 0.8$ & $ 
7.2\pm 0.1$ & $ -5.7\pm 0.5$ & $-16.1\pm 0.6$ & $\nu$ Sco B \\
3 & 396 & 196 & $<0.1 \%$ & $<0.1 \%$ & $ 240.8\pm 1.6$ & $ -22.5\pm 1.6$ & $ 7.1\pm 0.2$ & $  -7.9\pm  1.0$ & $  -16.1\pm 0.8$ & b Sco \\
4 & 156 & 67 & $<0.1 \%$ &$<0.1 \%$ & $ 241.6\pm 1.2$ & $ -20.5\pm 1.7$ & $ 6.5\pm 0.2$ & $  -7.1\pm  0.7$ & $  -16.2\pm 1.0$ & HD 144273 \\
5 & 166 & 44 & $12.2 \%$ & 0.2 \% & $  241.0\pm  2.2$ & $ -24.5\pm  1.8$ & $ 6.5\pm 0.2$ & $  -8.7\pm  1.1$ & $ -17.0\pm 1.0$ & HIP 77900 \\
6 & 58 & 13 & $ 13.5 \%$ &  18.4\% & $  241.8\pm  0.9$ & $ -19.0\pm  1.3$ & $  6.7\pm 0.2$ & $  -6.2\pm 0.5$ & $  -15.8\pm 0.9$ & HIP 78968 \\
7 & 45 & 23 & $ 1.0\%$ & $<0.1 \%$ & $  245.2\pm  0.7$ & $ -22.0\pm  0.8$ & $  7.4\pm 0.1$ & $ -7.3\pm  0.2$ & $  -16.1\pm 0.3$ & HIP 79910 \\
8 & 40 & 11 & $ 40.0 \%$ & $<0.1$ \% & $  244.8\pm  1.1$ & $ -22.6\pm  0.7$ & $  7.0\pm 0.2$ & $ -6.7\pm  0.7$ & $  -14.3\pm 0.6$ & HD 146457 \\
  \hline\hline
   \end{tabular}
   \label{tab:group_prop}
\end{table*}

\section{Age determination}
\label{section:age_determination}

The presence of a kinematic duality within Upper Scorpius poses new questions: is there any difference in the age distribution of the clustered and diffuse population? Is the distribution of $t_K$ reflected into an age spread between the groups? To provide tentative answers, we tried to compute the age of both the groups and the diffuse population in a threefold way: via usual isochrone fitting, through the fraction of disc-bearing stars and in a purely kinematic way.

The results of our analysis are summarized in Table~\ref{tab:results}.

\subsection{Kinematic age}
\label{section:kin_age}

We have seen in Fig.~\ref{fig:median_clustered} that at least part of the association was more clustered in the past, as visually evident in Fig.~\ref{fig:time_frames}. We might think to exploit this observation to constrain the age of the groups, since their very detection encapsulates, by definition, an age estimate: the moment of maximum spatial coherence might reflect that of the common birth of the members.

For each group, a quantitative estimate can be obtained by defining a coherence function $K(t)$ as the length of the minimum spanning tree connecting the points at time $t$. To minimize the impact of possible outliers, we excluded the $10\%$ longer branches for 2D estimates, and the $32\%$ for 3D estimates.

\begin{figure*}
\centering
\includegraphics[width=17cm]{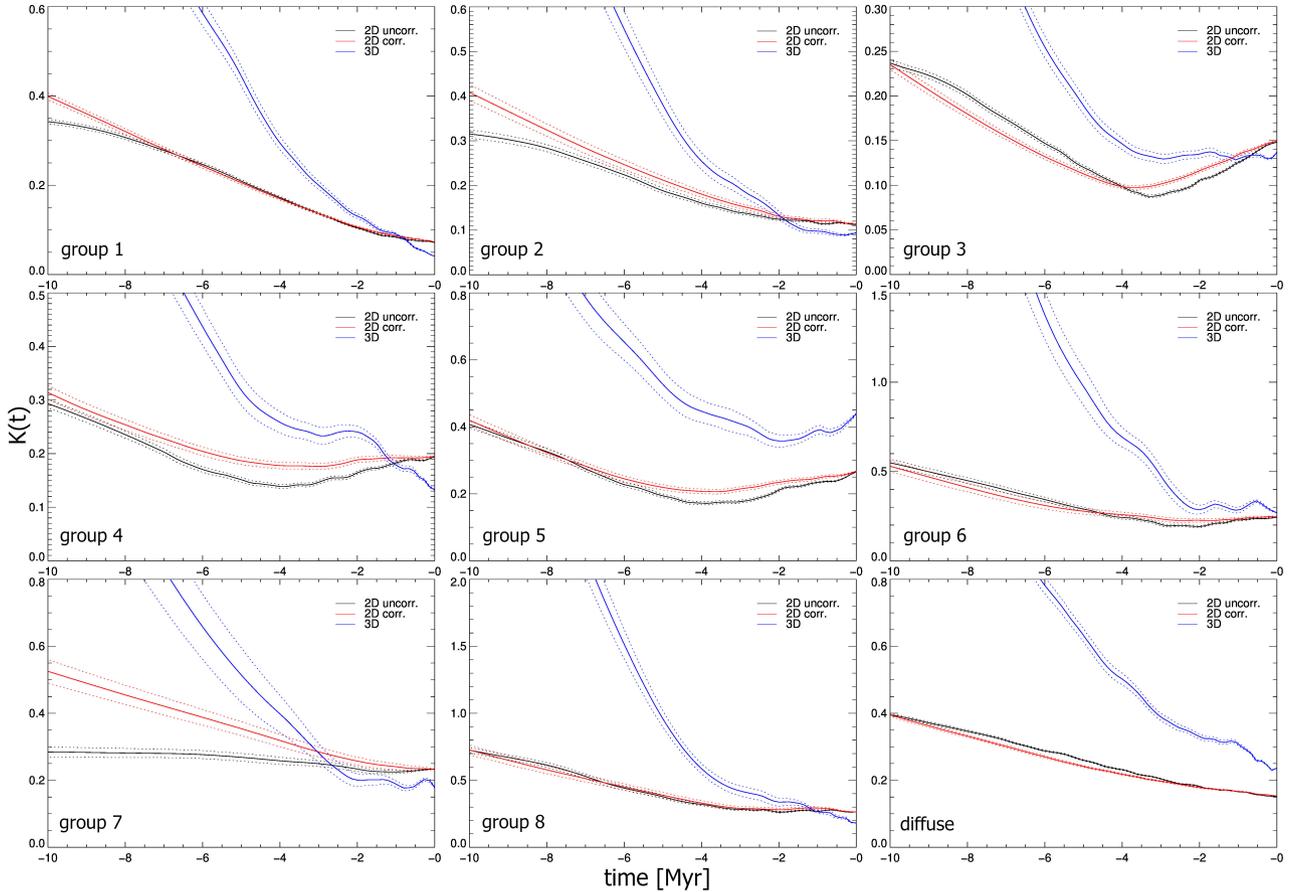}
\caption{Coherence function $K(t)$ for the groups, computed in the 2D case --with uncorrected proper motions (black) and corrected proper motions (red)-- and in the 3D case (blue), and shown in arbitrary units. 1$\sigma$ errors are shown as dashed lines. The minimum of $K(t)$ pinpoints a kinematic age. The diffuse population, as expected, does not cluster in the past.}
\label{fig:kin_age}
\end{figure*}

\begin{figure*}
\centering
\includegraphics[width=17cm]{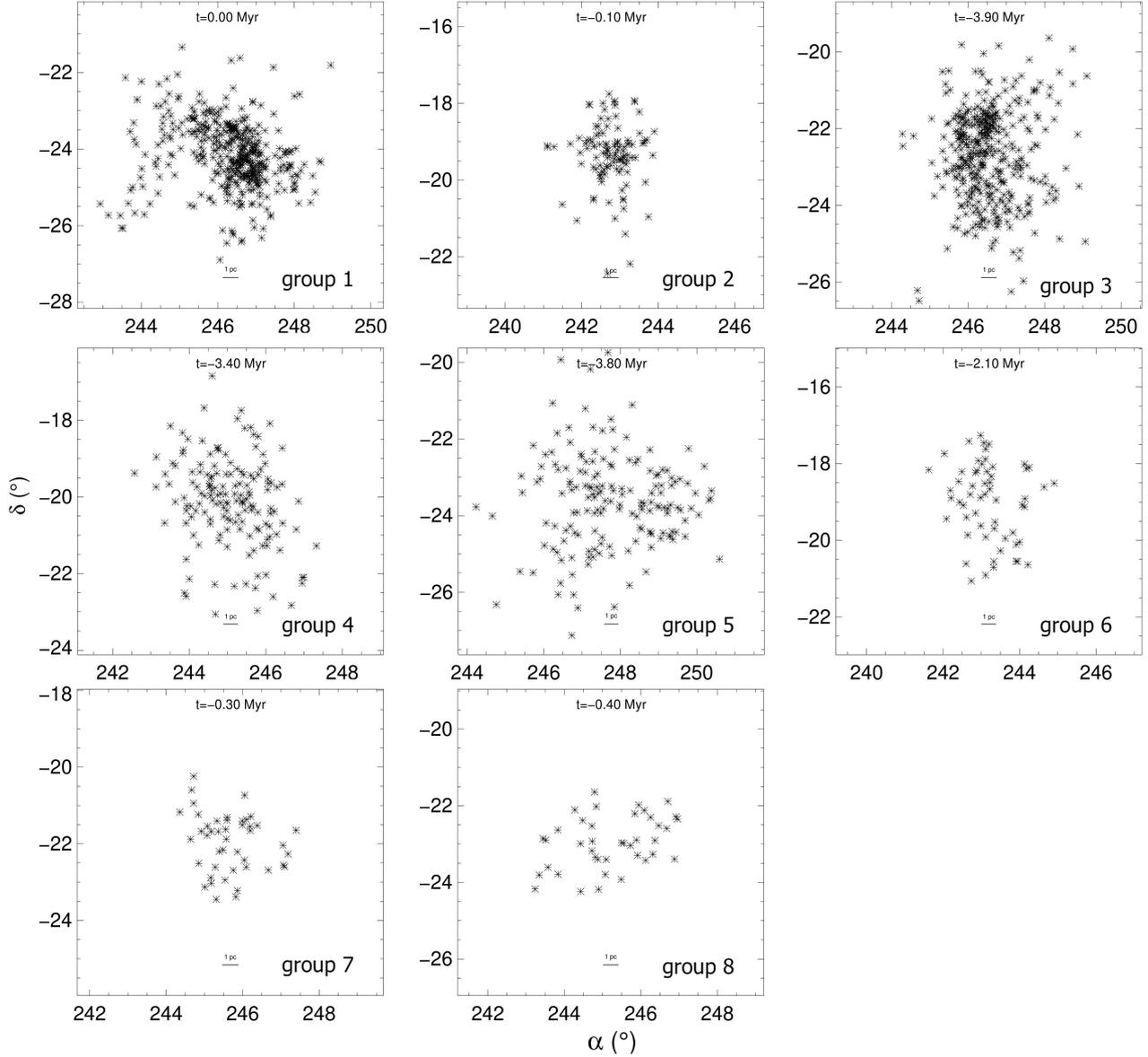}
\caption{All groups, at their maximum coherence, span just a few degrees in sky coordinates, corresponding to projected separations of a few pc.}
\label{fig:max_coherence}
\end{figure*}

The minimum of $K(t)$ provides us with an age estimate, that we will call {\it{kinematic}} age $t_K$. For 2D estimates we employed both corrected and uncorrected proper motion components, with minimal influence on the results: from this moment on, we will always think as $t_K$ as referring to the latter case. The trend of $K(t)$ for all the groups is shown in Fig.~\ref{fig:kin_age}, while the appearance of individual groups at $t=t_K$ is shown in Fig.~\ref{fig:max_coherence}. The error on $t_K$, again, was computed via $N=1000$ Monte Carlo simulations per group.

\subsection{Isochronal ages}
\label{section:iso_age}

We derive a second age estimate for the groups by employing the most common method of age determination for pre-MS stars: isochrone fitting.

We decided to employ BT-Settl CIFIST2011\_2015 isochrones \citep{2015aa...577A..42B} due to their large dynamical range ($m \in [0.01,1.4]$ $M_{\odot}$, $a \in [1,1000]$ Myr), spanning from young pre-MS objects in the brown dwarf regime to F-type stars. A constant solar metallicity, typical of most nearby star-forming regions, was assumed \citep{2011aa...526A.103D}.

The best-fitting solution for individual stars was found by leveraging as more as possible the available Gaia and 2MASS photometry: we averaged three estimates per stars, coming from the {\it{channels}} $\mathcal{F}=\{G,J\}$, $\mathcal{F}=\{G,H\}$ and $\mathcal{F}=\{G_{BP},G_{RP}\}$; in this way, it was possible to simultaneously employ the precision of Gaia $G$ magnitude, the long colour baseline of $G-J$\ and $G-H$, protecting against measurements errors, and a third channel that is simultaneously independent of $G$ and 2MASS data.

Whenever a filter in one channel did not fulfil the conditions described in Sect. ~\ref{section:photometry}, the corresponding channel was not used; also, a channel was ignored every time its total photometric error was larger than 0.15 mag; stars lying at more than $3 \sigma$ outside the region delimited by the isochrones were considered unfitted.

\begin{figure}
\centering
\includegraphics[width=9cm]{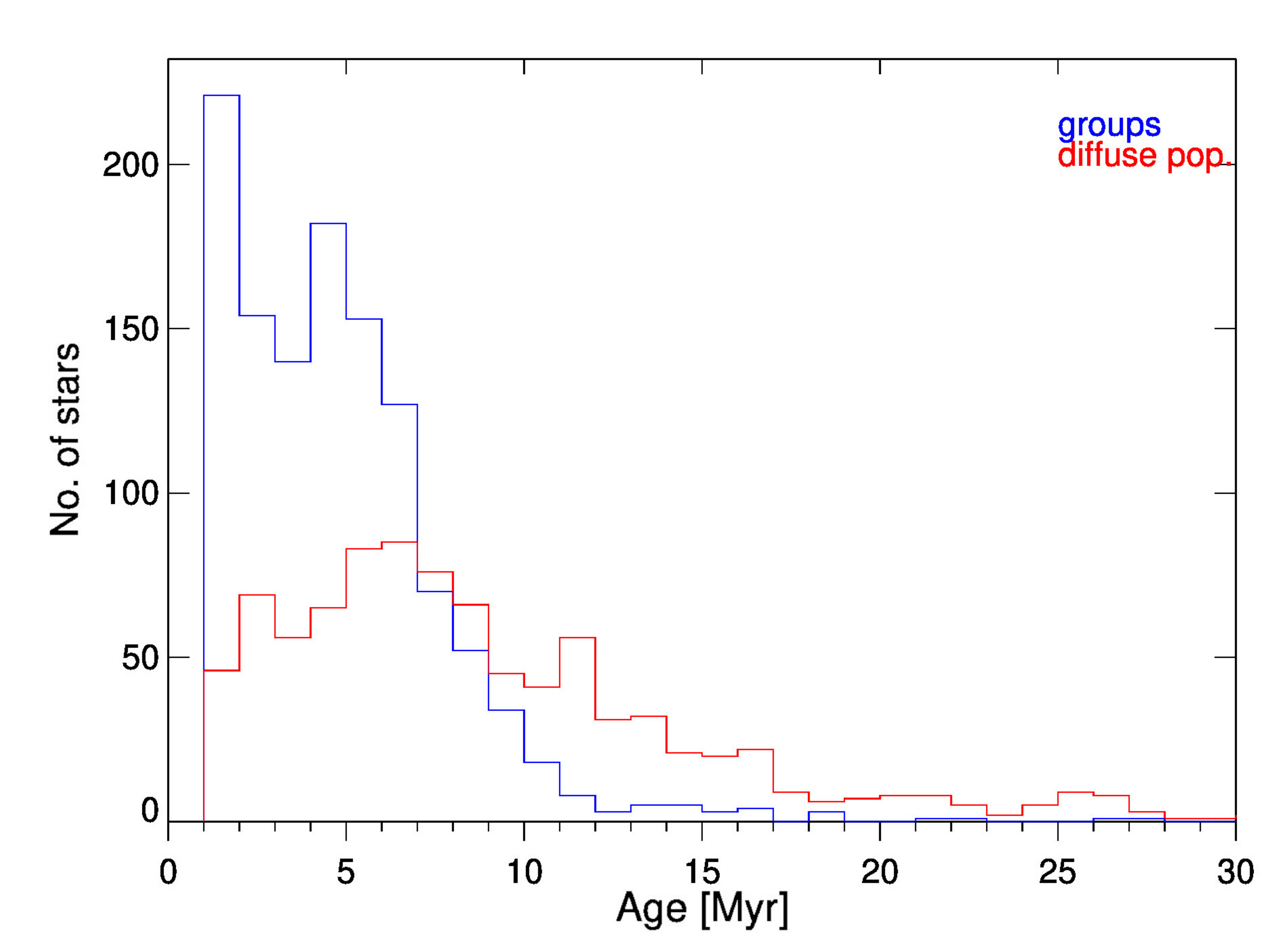}
\caption{Age distribution for the clustered (blue) and diffuse (red) population.}
\label{fig:groups_vs_diffuse}
\end{figure}

The results for the subgroups are shown in the second row of Table ~\ref{tab:results}: going from Group 1 to Group 7, the median age grows by $\sim 4$ Myr. What really stands out is the comparison between the clustered and the diffuse population (Fig.~\ref{fig:groups_vs_diffuse}): while the former has a median age of $\sim 4.5$ Myr and two well-defined peaks at $\sim 1$ Myr and $\sim 5$ Myr, the latter has a median age of $\sim 8.2$ Myr, with a much flatter distribution.

\subsection{Fraction of discs}
\label{section:disc_fraction}

Even though USCO has largely depleted its initial gas and dust reservoir \citep{2012ApJ...745...23M}, hints of accretion might still be found for some stars \citep{2009AJ....137.4024D}. Indeed, the disappearance of primordial discs, signalled by the opening of holes and gaps increasingly straining their initial SED \citep{2012ApJ...747..103E}, has been shown to occur with an exponential fashion \citep{2001ApJ...553L.153H}: the fraction of disc-bearing stars within young populations ($t \lesssim 10$ Myr) can therefore be used as an independent age estimator.

The long wavelengths of WISE {\it{W1, W2, W3, W4}} filters are particularly suitable for detecting young discs, with longer wavelengths probing larger distances from the star due to their intrinsic thermal structure \citep{1984ApJ...287..610L}. While IR excesses in {\it{W2}} and {\it{W3}} are suggestive of inner discs, {\it{W4}} is associated to a colder outer disc \citep{2018MNRAS.480.5099K}. The effect that we wish to look for is enormous: the luminosity of these discs in {\it{W3}} and {\it{W4}} can outshine that of the star itself \citep{2012ApJ...758...31L}. Previous works like \cite{2012ApJ...758...31L,2016MNRAS.461..794P} have greatly used {\it{W4}} magnitudes --often in combination with {\it{W3}}-- for this scope, due to its capability of probing outer disc zones.

To construct a reliable {\it{W3}} ({\it{W4}}) sample, we selected only data with the best photometric quality flag ('0'), photometric error $<0.2$ and apparent magnitudes $W3>3.8$ ($W4>-0.4$)\footnote{The last criterion, due to saturation problems leading to a flux overestimation \citep{2012wise.rept....1C}, was virtually unnecessary, given the distance of USCO.}. An insidious problem is that of {\it{flux contamination}}: since the angular resolution $\theta_i$ increases for redder bands\footnote{2'' for {\it{J}}, {\it{H}} and {\it{K}}, 6.1'' for {\it{W1}}, 6.4'' for {\it{W2}}, 6.5'' for {\it{W3}}, 12.0'' for {\it{W4}}.}, the measured flux becomes more at risk of including a non-negligible contribution of nearby, unresolved sources. To quantify this effect, all the sources within an angular distance $\theta_i$ were looked for in Gaia EDR3\footnote{Gaia EDR3 is essentially complete at $\theta>1.5''$, and can be relied upon until $\theta=0.7''$ for equal-mass sources \citep{2021aa...649A...5F}.}. Then, the $G$, $G_{BP}$, $G_{RP}$ fluxes were converted into $W3$ and $W4$ fluxes. A semi-empirical relation between input and output fluxes was used, combining the 8 Gyr BT-Settl isochrone for $0.01<M/M_\odot<1.4$ and the empirical tables by \cite{2013ApJS..208....9P} for $1.4<M/M_\odot \lesssim 20$.

The fluxes from field sources were derived averaging four estimates: from $G$ and parallax, if $\sigma_\pi/\pi < 0.5$, from $G$ and $G_{BP}$ if $\sigma_G+\sigma_{G_{BP}}<0.4$, from $G$ and $G_{RP}$ if $\sigma_G+\sigma_{G_{RP}}<0.4$, from $G_{BP}$ and $G_{RP}$ if $\sigma_{G_{RP}}+\sigma_{G_{BP}}<0.4$. Correction for extinction was applied whenever possible. If no estimate was available for even a single neighbour of a certain star, we employed $G$ fluxes, for consistency, for all the neighbourhood too.

The so-called contamination fraction $f_c$ was defined as $f_c=F_{cont}/F_{tot}$, where $F_{tot}$ is the flux from the $M$-band of interest and $F_{cont}$ the flux from all the field sources within $\theta_i$. Its effect consists in a magnitude decrease $\delta M=2.5 \cdot \text{log}_{10}(1-f_c)$. We subtracted it to measured magnitude: this correction goes {\it{against}} disc detection, as it makes magnitudes fainter. To be protected against the uncertainties in field star distances, ages, and on the correction itself, we decided to employ in this analysis only stars with $f_c<0.2$. This limits the corrections to $-\delta M \approx 0.25$ mag, far less than the expected effect for full discs (several mag).

The combination of quality and contamination cuts, though, greatly reduces the availability of $W4$ data, which additionally show a tendency for being {\it{always}} redder than expected, perhaps indicating debris discs \citep{2018AJ....156...71C} that are known to be common in the association \citep{2009ApJ...705.1646C}. For these reason, we decided at the end to employ only $W3$ magnitudes.

Abundant groups like 1 and 3 show a neat bimodality in their $(G-W3,G)$ CMD, with a second sequence of stars running parallel to discless stars, but with an offset of $\sim 2$ magnitudes. Hence, we chose to use the same criterion as \cite{2016MNRAS.461..794P}, i.e. we compute the excess $E(K-W3)$ relative to the expected colour, and identify full discs as those with $E(K-W3)>1.5$ mag. The choice of $K$ is based on the fact that its $\lambda$ ($\approx 2.2 \mu$m) is simultaneously too short to carry along a significant non-photospheric contribution and long enough to be protected against uncertainties in the extinction. We compute the expected $(K-W3)$ as that of the isochrone corresponding to the group isochronal age, computed in Sect.~\ref{section:iso_age}. We perform the same computation\footnote{Adopting the same threshold of 1.5 mag, as $G-$band emission is purely photospheric.} with $(G-W3)$, both as a consistency check and as a way to recover those few stars excluded by quality cuts in 2MASS photometry.

The fraction of primordial discs $f_D$ spans from $\sim 30 \%$ to $\sim 10 \%$, going from Group 1 to the diffuse population. Under the assumption of an exponential decay of $f_D$ \citep[see, e.g.,][]{2009AIPC.1158....3M}, we derived the {\it{disc age}}:
\begin{equation}
    t=-\tau \ln{f_D}
\end{equation}
where $\tau=2.5$ Myr \citep{2009AIPC.1158....3M}.

The fraction of discs we found within USCO is $f_d=0.19\pm 0.01$, comparable to that found by \cite{2018AJ....156...75E} and \cite{2020AJ....160...44L}.

The clustered population appears younger than the diffuse population, whose disc fraction has been computed using as expected colours those produced by individual ages and masses. Again, setting a fiducial line in this way produces a bias {\it{against}} disc detection and hence against an age spread \footnote{If the excess is computed starting from individual positions rather than from a fiducial line, it will be harder for the youngest stars to have an excess beyond 3$\sigma$; the opposite applies for stars older than the fiducial line, but they are usually not expected to show an IR excess at all. The overall effect goes against disc detection.}.

A comparison of our sample of disc-bearing stars with that by \cite{2020AJ....160...44L} shows that, among 161 full discs having a cross-match, 145 (148) are there identified as full-discs, the remaining 16 (13) being labelled as debris/evolved transitional discs, for the criterion employing $G-W3$ ($K-W3$). It is significant that the number of false negatives, i.e. sources not labelled here as full discs but that are identified there as such, is comparable: 84/1261 (81/1261). We can reasonably assert that our disc fraction estimates are not biased in one direction or another.

\begin{figure}
\centering
\includegraphics[width=9cm]{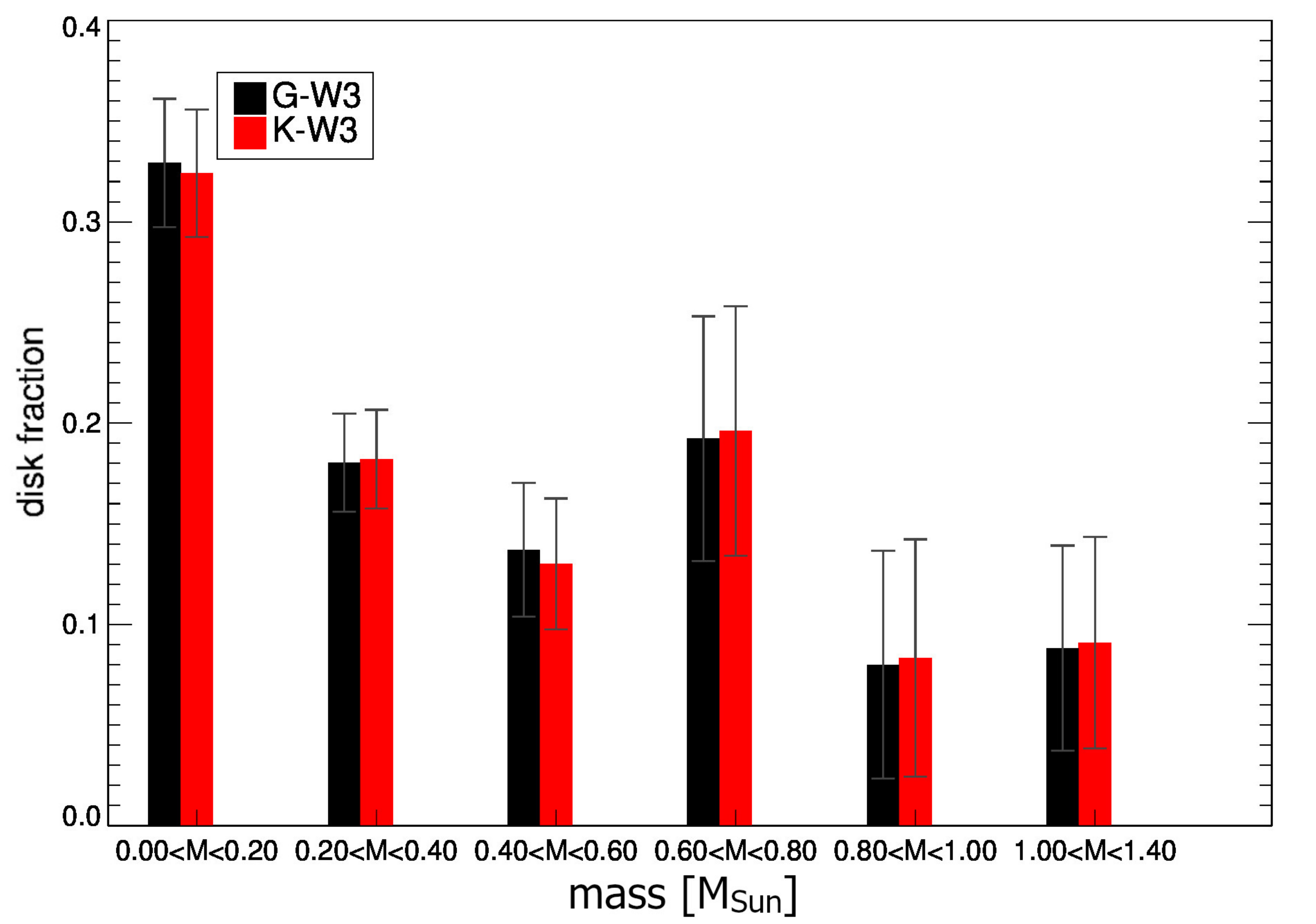}
\caption{Disc fractions for different bins of stellar mass. Black bars give the fraction of stars where disks were detected from the G-W3 colour;
red bars those with disks detected from the K-W3 colour. Going towards bins of increasing stellar mass, a clear decrease in the disc fraction is seen.}
\label{fig:disc_mass_bias}
\end{figure}

The assumption of a single $\tau$ hides the different timescales of disc decay with stellar mass: it has been shown that the lifetime of a disc steadily decreases with stellar mass \citep{2015aa...576A..52R}. The dependence of the disc fraction on stellar mass is shown in Fig.~\ref{fig:disc_mass_bias}: the fractions for the first bin and the last two are in complete agreement with the dedicated IR survey performed by \cite{2012ApJ...758...31L} and with \cite{2016MNRAS.461..794P}.

\begin{table}
  \centering
   \caption{Age estimates obtained through photometry ($t_P$), discs ($t_D$) and kinematics ($t_K$). The number of stars in each group ($n$), the sample standard deviation of isochronal ages ($s$) and the fraction of discs ($f_D)$ are shown too.}
   \begin{tabular}{lcccccc} \hline \hline
Group & n & $t_P$ [Myr] & $s$ & $f_D$ (\%) & $t_D$ [Myr]& $t_K$ [Myr] \\
            \hline
1 & 467 & $2.6 \pm 0.1$ & 1.5 & $31\pm 3$ & $3.0 \pm 0.3$  & $0.0 \pm 0.1$ \\
2 & 114 & $4.4 \pm 0.1$ & 1.0& $27\pm 6$ &  $3.2_{-0.5}^{+0.6}$  & $0.1 \pm 0.2$ \\
3 & 396 & $4.8 \pm 0.1$ & 1.1& $25\pm 3$ &  $3.5 \pm 0.3$  & $3.7 \pm 0.3$ \\
4 & 156 & $5.2 \pm 0.1$ & 1.4& $17\pm 4$ &  $4.5_{-0.5}^{+0.7}$  & $3.4 \pm 0.6$ \\
5 & 166 & $6.1 \pm 0.2$ & 1.6& $14\pm 4$ &  $4.8_{-0.6}^{+0.8}$  & $3.8 \pm 0.4$ \\
6 & 58 & $4.4\pm 0.2$ & 1.2& $26\pm 8$ &  $3.4_{-0.7}^{+0.9}$  & $2.1 \pm 0.6$ \\
7 & 45 & $6.2 \pm 0.2$ & 1.2& $9\pm 4$ &  $6.0_{-1.0}^{+1.7}$ & $0.3 \pm 0.4$ \\
8 & 40 & $5.4 \pm 0.3 $ & 1.7& $12\pm 6$ &  $5.3_{-1.0}^{+1.7}$ &  $0.4 \pm 0.8$ \\
  \hline
clust. & 1442 & $4.5\pm0.1$ & 1.5 & $24\pm2$ & $3.6\pm0.2$ & \textemdash \\
diff. & 1303 & $8.2\pm 0.1$ & 3.7 & $10\pm1$ & $5.7\pm0.3$ & \textemdash \\
  \hline\hline
   \end{tabular}
   \label{tab:results}
\end{table}

\begin{figure}
\centering
\includegraphics[width=9cm]{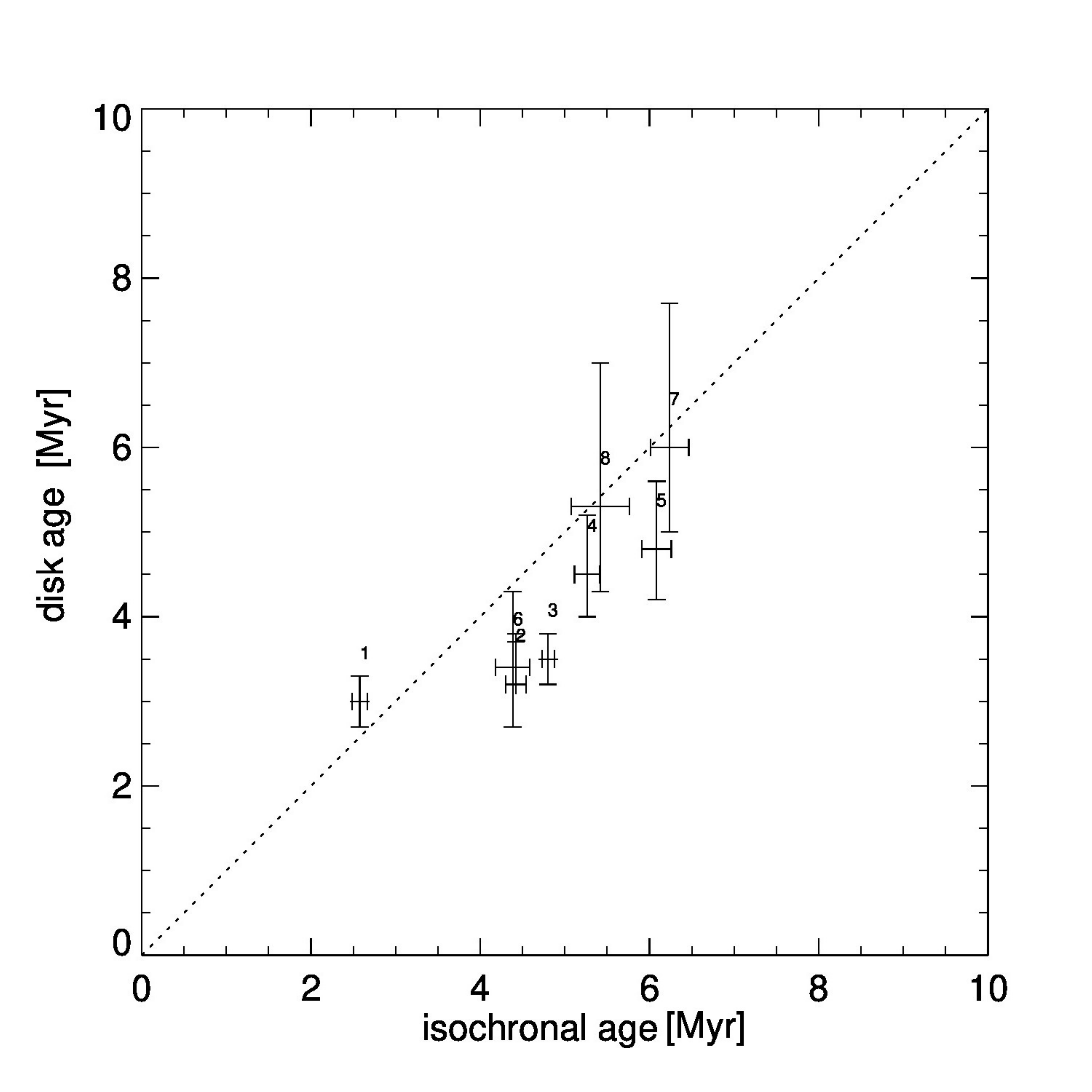}
\caption{Isochronal vs disc ages for USCO groups. These estimates show a remarkably high correlation (0.84).}
\label{fig:age_comp1}
\end{figure}

\begin{figure}
\centering
\includegraphics[width=9cm]{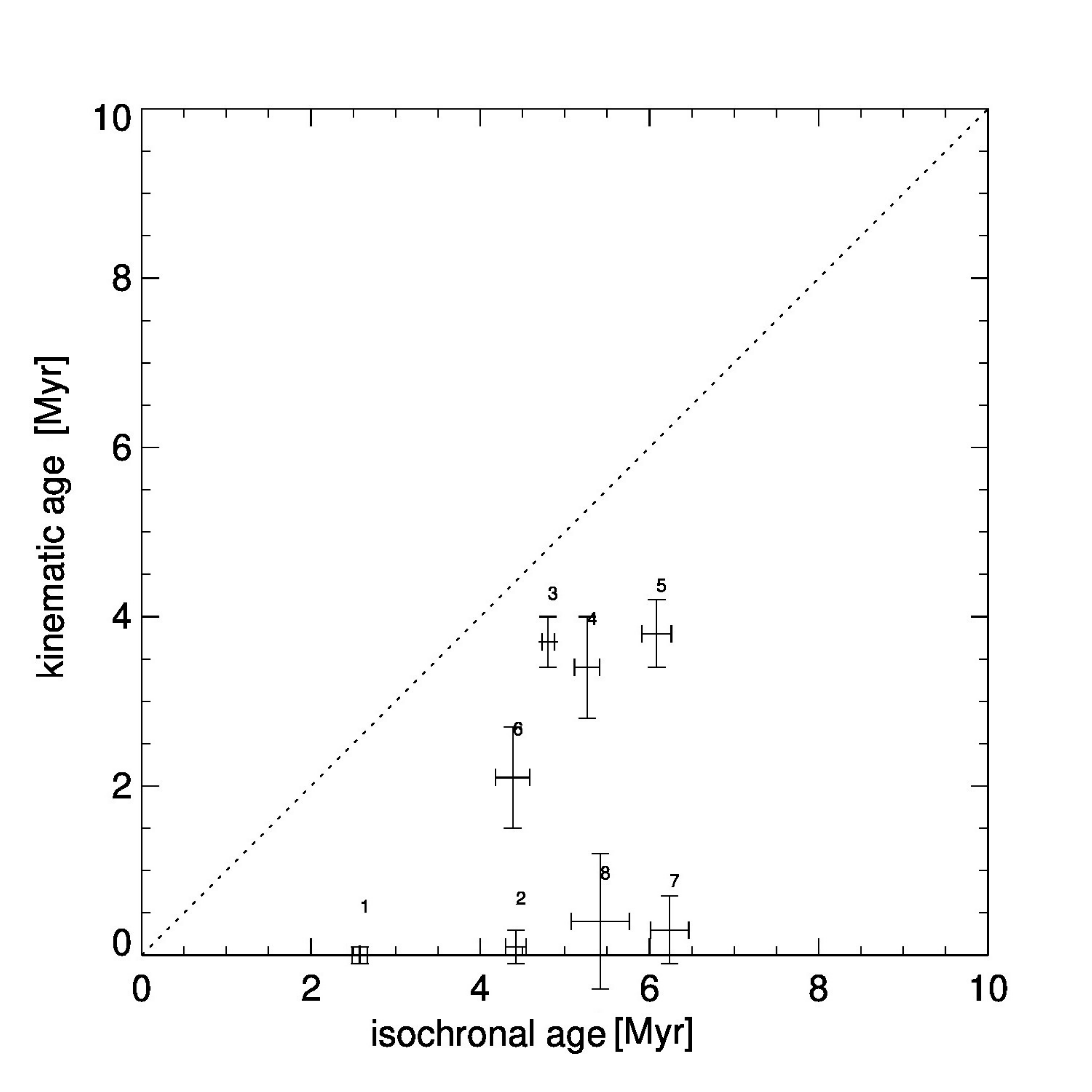}
\caption{Isochronal vs kinematic ages for USCO groups. Kinematic ages always appear to be underestimated.}
\label{fig:age_comp2}
\end{figure}

\begin{figure}
\centering
\includegraphics[width=9cm]{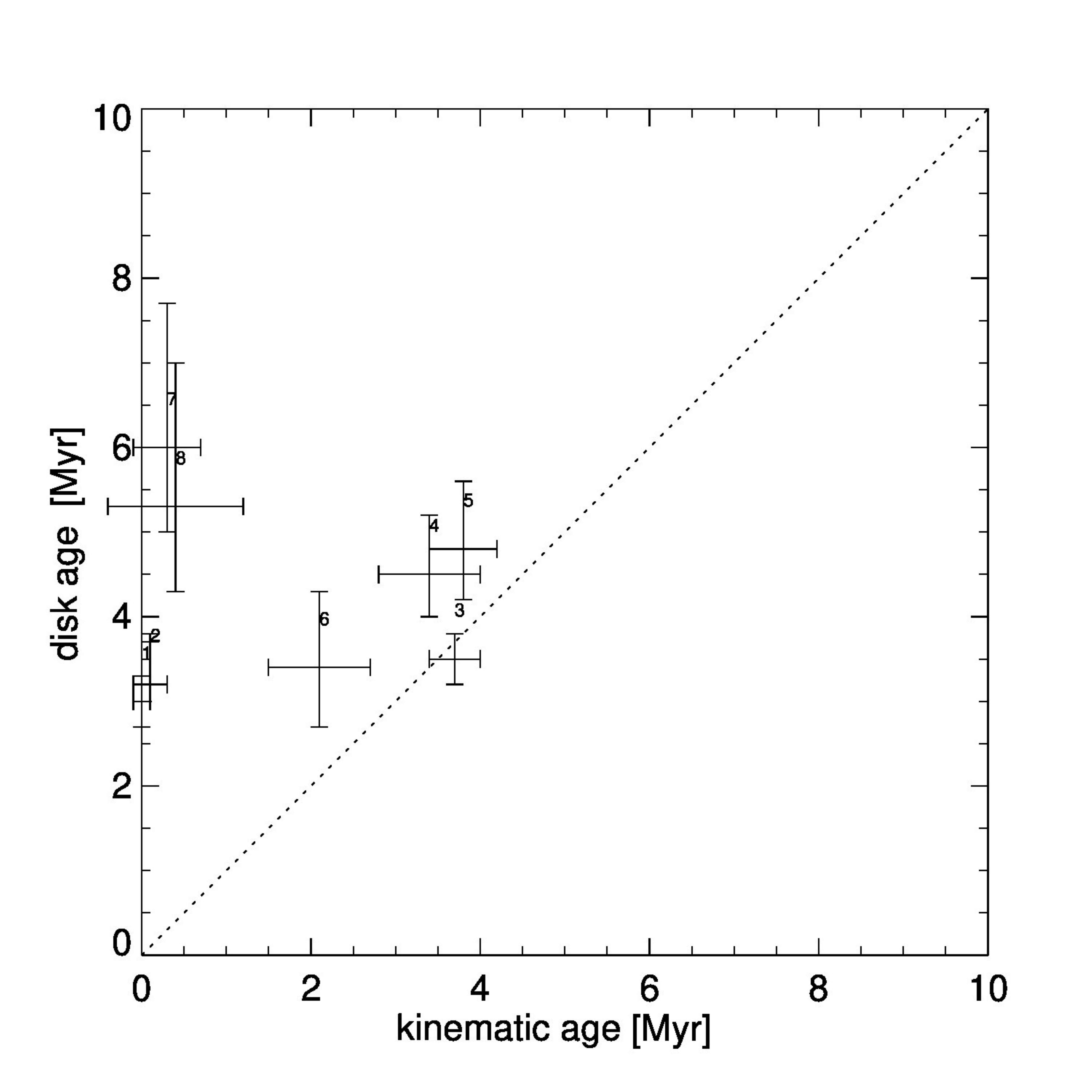}
\caption{Kinematic vs disc ages for USCO groups. The same considerations as in Fig. ~\ref{fig:age_comp2} apply.}
\label{fig:age_comp3}
\end{figure}

We show in Fig.~\ref{fig:age_comp1},~\ref{fig:age_comp2},~\ref{fig:age_comp3} the comparison between the three estimates for the eight groups. While disc and isochronal age show a remarkable correlation (0.84), kinematic ages appears always underestimated, hinting at the presence of factors not taken into account so far. We will discuss the findings of the age analysis in Sect.~\ref{section:discussion}. 

\section{Discussion}
\label{section:discussion}

\subsection{Kinematic analysis}
As many recent studies have shown \citep[e.g.,][]{2020aa...638A..85R}, the formation of associations cannot be reduced to the simple scenario of a monolithic burst. The idea of a coherent, uniform expansion has given way to that of a complex star formation history, prolonged over million of years and with a large spatial variability. While this might seem an argument against kinematic reconstruction \citep[e.g.,][]{2016MNRAS.461..794P}, the substructures themselves dissolve over time \citep{2019aa...623A.112D}, so they might in principle be targeted by the same approach.

The presence of a high degree of substructure within Upper Scorpius has been recently quantified by \cite{2021aa...647A..14G}, confirming a long-held suspicion \citep[e.g.,][]{2018MNRAS.476..381W}. However, \cite{2019aa...623A.112D} concluded that this complex structure seen at present time cannot be unfolded and brought out into a global pattern of motion: regions of higher and lower density do not bear distinct kinematic imprints, and close proximity on the sky is not equivalent to proximity in velocity, as if the different populations are mixed up. This is similar to what is observed in young clusters \citep[e.g.,][]{2014aa...563A..94J,2015AJ....149..119T,2015aa...574L...7S,2017aa...602L...1D}. The general trend that denser regions appear younger than the sparser ones hints at the presence of multiple populations or, at least, of an extended star formation history.

The initial substructure of a group of stars can be irremediably altered not only by the violent relaxation following the removal of the initial gas reservoir \citep{10.1093/mnras/136.1.101}, but also by subsequent dynamical interactions \citep{2002MNRAS.334..156S}, leading to its complete disappearance, if the dynamical timescale of the region is shorter than its age\footnote{The picture is even more complex, as the dynamical timescale itself can change over time as a result of its density variations, with the counter-intuitive result that the presence of high degree of structure today rules out the existence of highly compact states in the past \citep{2014MNRAS.445.4037P}.}. If, like in associations, dynamical interactions may be neglected, a memory of the initial velocity pattern can be preserved for a longer time \citep[][]{2004aa...413..929G}, since the main factor leading to the erosion of the original velocity structure now becomes the galactic tidal field, acting on timescales of $\sim 10^7$ yr \citep{2018MNRAS.476..381W}.

\cite{2018MNRAS.477L..50G}, employing Gaia DR1 data of 1322 stars, found that the shape of USCO is approximately ellipsoidal; they conclude that the association is not dynamically relaxed, meaning that its shape is an imprint of its star formation history. Based on these considerations, we decided to employ a kinematic approach to see how much the initial structure of the association is still perceptible beneath its present velocity structure.

The method employed was the classical linear trace-back, in which individual star motions are traced back in time. The approach differed from the classical studies of USCO both in depth, number of stars and purposes: contrary to earlier studies, with typical sample sizes of a few hundreds, our sample comprises about 3000 stars and did not aim at finding a single kinematic age for the whole USCO --something already excluded by \cite{2012ApJ...746..154P}--, but rather at investigating its degree of substructure.

As can be imagined, the feasibility of a trace-back analysis is ultimately limited by the precision of the available velocities \citep{Goldman:2018IAUS}. The need for precise radial velocities \citep[$\sigma_v << 1$ km/s,][]{2017ApJ...850...11D} was here satisfied by combining RV data from APOGEE and GALAH to those provided by Gaia: we have built a catalogue of 771 USCO sources (our "3D sample") with median velocity uncertainty $\sigma_v=0.12 \text{ km s}^{-1}$. Being the 3D sample size only $\sim 30\%$ of the 2D sample, the former was employed to verify and complement the results of the main analysis, carried in 2D.

Our cluster analysis found eight groups, which are seen to clump at different times at the past. The real nature of Groups 1, 2, 3, 4, 5 and 7 is extremely likely, comparing it with random clustering of sources in 2D or in 3D, while Groups 6 and 8 would need additional RV measurements to confirm their physical nature. None the less, it is worth mentioning that the disc fraction of Group 6 is higher than that of the field with a confidence level of $\sim 2\sigma$.

The results for Group 1, that should be interpreted as $\rho$ Ophiuchi's bark, the region that is not too extinct to be undetectable in Gaia's optical bands, are consistent with the study by \cite{2011AJ....142..140E}, that derived a disc fraction of $27 \pm 5\%$ and an isochronal age of about 3.1 Myr. When cross-matching our USCO sample with the $\rho$ Ophiuchi sample by \cite{2019aa...626A..80C}, we find that 5 \% of its sources belong to our Group 7 and another 5 \% to our Group 8. The subdivision of sources between Group 0 and Group 8 closely resembles that of \cite{2021arXiv210112200G}, with compatible ($\alpha$, $\delta$, $v_a$, $v_d$, $\pi$) between our Group 1 and their "Pop 1" and our Group 8 and their "Pop 2". The concentration of sources in the northern region of USCO ($352^\circ<l<355^\circ$, $22^\circ<b<25^\circ$), coming with a complex velocity structure, was already noted by previous studies \citep{2019aa...623A.112D,2021aa...647A..14G}; here we suggest the division of those stars in three different groups (2, 4 and 6) as a possible solution to this conundrum.

A particular fruitful comparison can be done with the recent work by \cite{2021arXiv210509338K}, which employs a density-based hierarchical clustering algorithm to identify substructures within young nearby associations. Contrary to their approach, based on a thorough inspection of the present phase space structure, our simpler, semi-automated method tried to incorporate in the classification scheme the time variable too. This is what led us to identify, for instance, the rather spread and populous Group 3, which was very concentrated in the past. A comparison of their Table 6 with our Table~\ref{tab:group_prop} likely leads to associate their Group E, F, G, H and I with our Groups 2, 7, 6, 3, 1, respectively.

The most interesting result came when putting together the groups, and comparing them with the stars that could not be put in a group. The clustered population appears to have been more compact in the past (with a peak at $t \sim 3.5$ Myr ago, dominated by Group 3), with a confidence level on this result of $14 \sigma$. This is not true for the diffuse population, which instead shows its tightest configuration at present time. These results came as a surprise, as the age for USCO members should be 5-11 Myr, and led us to investigate whether the retrieved substructures were correlated to age gradients within the region: to this aim, we employed both isochrones for pre-MS and discs.

\subsection{The age conundrum}

The first robust result of the age analysis is that the clustered population appears younger than the diffuse population, in a similar way as \cite{2019aa...623A.112D} noticed with their distinction based on the present projected source density. The natural explanation for this fact is that star formation in USCO appears to have happened in small groups that disperse after a few Myr, dissolving in the field of the older population, but retaining for some time memory of their original velocity structure.

The clustered population itself shows an internal age gradient, as expected from previous works on the region \citep{2016MNRAS.461..794P}. The youngest group is Group 1 ($\rho$ Ophiuchi), with a $t \sim 3$ Myr consistent with the literature \citep{2011AJ....142..140E}; Group 2, 3 and 6 are approximately coeval, while the small comoving Group 7 appears the remnant of an older formation event.

An independent age estimate was obtained through the disc fraction $f_D$, defined as the fraction of stars within the sample still bearing marks of a primordial disc. We decided to follow the criterion for distinction of disc classes used by other studies \citep{2012ApJ...758...31L,2016MNRAS.461..794P}, but the enormous restriction of W4 data, due to a combination of quality cuts and contamination from the field, made us lean towards the use of the sole criterion on W3. We employed as colour benchmark both K-W3 and G-W3, finding consistent results, and validated our proxy with a cross-match with \cite{2020AJ....160...44L}. The price to pay was the impossibility to reliably identify looser disc evolutionary stages (evolved, transitional), which none the less are not used when inferring the age of a population. 

We find an average full disc fraction of about $0.19\pm 0.01$, consistent with previous work \citep{2006ApJ...651L..49C,2012ApJ...758...31L,2020AJ....160...44L,2018AJ....156...71C}, and a well-defined trend with stellar mass consistent with \cite{2012ApJ...758...31L}. The disc fraction is again significantly higher for the clustered ($0.24\pm0.02$) than for the diffuse population ($0.10\pm0.01$). Since disc statistics has been shown not to be influenced by binarity, as circumbinary discs in USCO decay with the same pace as circumstellar discs \citep{2018MNRAS.480.5099K}, the robust correlation of $f_D$ (0.84) with isochronal estimates reinforce the idea of a real age spread between the groups.

It should be underlined that, whilst recognising full disc can be considered safe and little model-dependent, the e-folding time of disc decay is usually derived from comparison with isochronal ages, generating a circular argument when confronting $t_K$ and $t_P$. Estimates of $\tau$ include 2.3 Myr \citep{2010aa...510A..72F}, 2.5 Myr \citep{2009AIPC.1158....3M}, 3 Myr \citep{2014aa...561A..54R} and even 
5 Myr \citep{2018MNRAS.477.5191R} for models including magnetic-driven radius inflation; additionally, $\tau$ has been shown to depend on environmental conditions like the mass and the rotation rate of the parent molecular cloud core, and on the stellar mass itself \citep[e.g.,][]{2006AJ....131.1574L, 2009AIPC.1158....3M,2015aa...576A..52R}. Therefore, our disc ages must be considered relative, rather than absolute.

The issue with isochronal ages is similar and, if anything, even more severe, as the history of age determinations in USCO bears witness. The first systematic studies of the region argued for a uniform age of $\sim 5$ Myr \citep{1999AJ....117.2381P,2002AJ....124..404P}, but then \cite{2008ASPC..384..200H} showed that the ages of young stars depend on the spectral class, with low-mass star ages undestimated by 30-100\% and high-mass star ages overestimated by 20-100\%; a twofold variation of the inferred ages for intermediate and low-mass stars emerged for other clusters too \citep{2008MNRAS.386..261M,2009MNRAS.399..432N,2013MNRAS.434..806B}. Significant insights on the problem are given by binary systems for which dynamical mass estimates are available: discrepancies between ages inferred for presumably coeval components have been confirmed by \cite{2016ApJ...817..164R} and \cite{2019aa...622A..42A}. A total reassessment of the problem of age determination in USCO was carried out by \cite{2012ApJ...746..154P}, who used F-type stars as a benchmark to establish a revised age of $11\pm 2$ Myr. The picture was confirmed in a subsequent study \citep{2016MNRAS.461..794P}, alongside with a strong dependence of age estimates on spectral class, with younger ages for both K- ($5\pm2$ Myr) and B-type stars ($7\pm2$ Myr) than for F- ($10\pm 1$ Myr) and G-type ($13\pm 1$  Myr)  stars. Interestingly enough, the derived age spread for the association was as large as 7 Myr.

While the debate on the age of USCO is still ongoing \citep[e.g.,][]{2019ApJ...872..161D}, a parallel discussion involves the above-mentioned age spread, either attributed to an extended star formation or to systematic effects inherent to models. An interesting solution has been put forward by \cite{2016aa...593A..99F}: highlighting the difficulties in modelling convection for pre-MS stars, he noticed that the effect of magnetic fields on a protostar --stronger for less massive stars-- is to slow down its radial contraction along the Hayashi line; the resulting luminosity at a fixed age is higher than predicted, leading to incorrectly infer younger ages if the effect is not taken into account. By means of Dartmouth magnetic models \citep{2012ApJ...761...30F} equipped with the maximum allowed surface magnetic field strength, he found that a consistent 10 Myr isochrone could fit the observed Hertzsprung--Russell diagram (HRD) of USCO across all spectral types. The explanation by \cite{2016aa...593A..99F}, though, can create problems with moving groups with well-defined age \citep[Tucana-Horologium association, $\beta$ Pictoris; see][]{2015MNRAS.454..593B}.

A completely different approach is that of \cite{2017ApJ...842..123F}, who emphasize the importance of incorporating the star formation history (SFH) into the study of associations. Dividing stars by spectral type is risky because, at fixed mass, a younger star is of a later type than an older one.  While low-mass stars ($M<0.4 M_\odot$) enter the MS via the Hayashi line, intermediate stars develop a radiative core that makes them move along the horizontal Henyen track; given that the ascension toward earlier spectral types acts on a timescale of 1-10 Myr, a population forming through a burst of comparable duration will generate spurious age differences if divided according to spectral class.

We will discuss how our results can be reconciled with this framework in Sect. ~\ref{section:sfh}, shifting for the moment the focus to the discussion of some biases impacting on both kinematic and photometric considerations.

\subsection{Assessment of biases}

A particularly tricky bias, affecting both the kinematics and isochronal ages, is that resulting from unresolved binaries. Assessing the fraction of binaries $f_b$ in stellar populations has been the aim of several studies in the last years. A general trend is the increase of $f_b$ with the primary mass:  $f_b = 27\pm 1 \%$ for $M=0.075-0.6 M_\odot$ \citep{2019AJ....157..216W},  $f_b = 41\pm 3 \%$ for $M= 0.78-1.02 M_\odot$ \citep{2010ApJS..190....1R}, $f_b = 54\pm 4$\% for $M= 1.02-1.25 M_\odot$ \citep{2010ApJS..190....1R}; however, environmental effects can play an important role, as shown by USCO itself that, contrary to the field, shows a fairly constant $f_b=35 \%$ for early M and G-K stars \citep{2020AJ....159...15T} and a value as high as $\sim 70\%$ for B-A-F stars \citep{2005aa...430..137K,2007aa...474...77K}.

We envisage the effect of unresolved binary stars on our sample to be twofold. As regards the kinematics, we expect that the wobbling of the photocentre is reflected in a perturbation of the system's true proper motion with the result that, when performing kinematic trace-back, binaries will be preferentially directed towards the diffuse population rather than to the clustered population. As regards the photometry, the presence of a cooler secondary star can shift the whole system towards lower temperatures and higher luminosities in the HRD. The combination of the two effects makes unresolved binaries appear younger than they are. It has recently been shown \citep{2021arXiv210309840S} that this factor is able to create a large apparent spread in a coeval population, especially when accompanied by high parallax uncertainties. None the less, the correlation that we found between the isochronal age and the disc age is reassuring, as the latter is expected to be independent of multiplicity \citep{2018MNRAS.480.5099K}.

To quantify the impact of these considerations on our results, we started from the sample of 614 USCO stars constructed by \cite{2020AJ....159...15T} to assess the multiplicity in the region. Their sample, covering the mass range $[0.7,1.5] M_\odot]$, is invaluable for two reasons: on the one hand, it was accurately vetted not to be biased against or towards binaries; on the other hand, it was extensively studied to look for companions by means of speckle interferometry, pushing the detection sensitivity well below Gaia's\footnote{The detection limits of their survey, outside $\theta \approx 0.1''$ (corresponding to a projected physical distance $d \approx 15 AU)$, are such that the survey is complete at a contrast $\Delta I=2$ mag.}.

A tentative diagnostic of possible binarity is the renormalised unit weight error (RUWE), a parameter associated to each source of Gaia EDR3 and quantifying how much the assumption of {\it{isolated point source}} is suited to the astrometric solution; values of RUWE>1.4 are usually used as a threshold to flag a potential non-single star nature \citep{2021aa...649A...5F}. After removing 70 companions of their list that are resolved by Gaia (having angular separations $\theta \gtrsim 1''$), we are left with a sample of 180 companions. We have two samples of primary stars, that we may dub "single" (S) and "binary" (B). We retrieved the RUWE parameter for all these stars and verified that the fraction of stars with RUWE>1.4 ($f_R$) is significantly higher for sample B ($76\%$) than for sample S ($19\%$). On the ground of the expected multiplicity of USCO, we would expect for our 2D sample a $f_R \sim 30-35\%$. Instead, we find that $f_R=16 \pm 1\%$, equal for the clustered ($15 \pm 1\%$) and the diffuse ($17\pm 1 \%$) populations. Exceeding the natural bias against strict binaries proper to Gaia-based samples \citep{2020AJ....159...15T}, our selection criteria appear to have preferentially excluded binaries from the 2D sample.
 
To verify it explicitly, we applied the same selection criteria of Table~\ref{tab:criteria}. Whereas we recover $87\%$ of stars from sample S, we only recover $67\%$ from sample B. Although the median parallax uncertainty is larger (0.089 vs 0.023 mas), the effect is mostly due to velocity cuts. This basically means that proper motions of unresolved binaries can be so large to slide them out of the selection window $25\%$ of the time.

The high-RUWE stars of our clustered (diffuse) sample are significantly younger than their parent population, having a median $t=2.6$ Myr ($t=4.4$ Myr). To verify if this result is consistent with a population of unresolved binaries, we set up a simulation of 10000 binary systems, with initial mass function (IMF) and mass ratio distribution (CMRD) as in \cite{2011ApJ...738...60R}. After randomly generating primary masses according to the IMF and secondary masses according to the CMRD and the IMF, we assigned each star a set of magnitudes ($J$, $H$, $G$, $G_{BP}$, $G_{RP}$), coming from the same set of models used in the isochronal analysis, and a fixed age. For an age of $t=5$ Myr (comparable to the BT-Settl result for the clustered population), the derived median age shifts to $t=3.0$ Myr; for $t=8$ Myr (similar to the diffuse population), it shifts to $t=4.8$ Myr. The similar relative magnitude of the age deviation, coupled with the comparable fraction of high-RUWE stars, does not constitute an argument against the age spread between the two kinematic populations of Upper Scorpius, but naturally explains the young tail seen in the diffuse one (Fig.~\ref{fig:groups_vs_diffuse}).

\begin{figure}
\centering
\includegraphics[width=9cm]{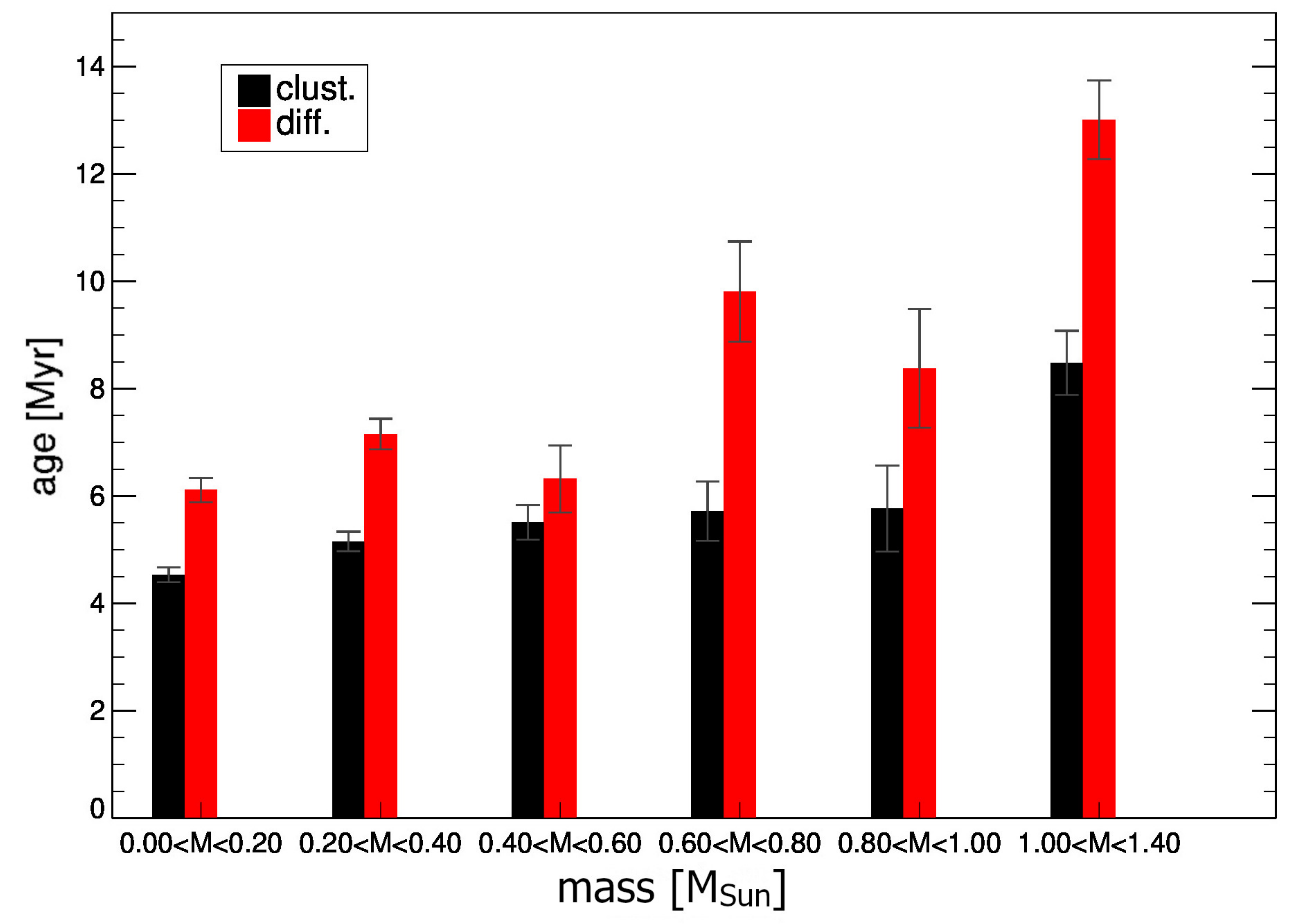}
\caption{Average isochronal ages for different bins of stellar mass. Black bars give the average age for members of the clustered population;
red bars for those of the diffuse population.}
\label{fig:age_mass_bias}
\end{figure}

A second issue worth considering is the already mentioned difficulty in assessing ages of low-mass stars. If we divide our sample in bins of fitted mass, we recover 119 stars with $M>1 M_\odot$\footnote{The smallness of the sample is due not only to the IMF but also to selection effects: bright stars at the typical distance of USCO tend to have poorer astrometric solutions than low-mass stars.}): 80 stars in the diffuse, 39 in the clustered population, with median ages 13.7 Myr and 8.1 Myr, respectively (Fig.~\ref{fig:age_mass_bias}). The unequal division of mass might point to different properties of the two populations; \cite{2018MNRAS.477L..50G} presented evidence for a somewhat different spatial distribution of the brightest and the faintest stars in USCO, hint of a different relaxation state that would imply an earlier formation of massive stars. We don't have the means to answer this question, but even a general increase of our fitted ages would not be detrimental to our main arguments.

As concerns field contamination, expected to some extent due to the rough definition of the 2D sample, it tends to create an age overestimate. Starting from the 3D sample, we applied an iterative exclusion of the stars that are more than 3 sigma out of the velocity distribution, up to convergence: 82/924 stars ($\sim 9\%$) are excluded in this way. The median of their ages is twice as large (11 Myr) as that of the full sample. A similar fraction of interlopers is found by applying the same algorithm to the age distribution of the 2D sample: out of 2183 stars with an age estimate, those excluded are 210 ($\sim 10\%$), mostly (91\%) belonging to the diffuse group.

\subsection{The star formation history of Upper Scorpius}
\label{section:sfh}

Turning back the attention to our main point, we might notice that the fraction of stars found in the clustered population is $\sim 50\%$. This fraction, likely underestimated because of field contamination, is much higher than the $14.5 \%$ found by \cite{2019aa...623A.112D} across the whole Sco-Cen, suggesting an evolutionary scenario of kinematic structures, consistent with the younger age of USCO compared to UCL and LCC. In the former, even though the present space distribution is quite well-mixed, the velocity distribution is not, and still allows a distinction of subgroups; but velocity patterns are beginning to fade away, too, as confirmed by the diffuse population, that has lost, or is losing, its initial kinematic imprint.

The classical picture of star formation in Scorpius-Centaurus, first put forward by \cite{1999AJ....117.2381P}, envisions the first outburst of star formation \citep[not before 30 Myr, see][]{2019aa...623A.112D} in LCC, that triggered shortly after activity in the adjacent UCL, and finally events like the explosion of supernovae led to star formation in USCO. But this should be considered no more than a zeroth order approximation of a more complex sequence of events. Indeed, a thorough study by \cite{2016MNRAS.461..794P} proved the existence of an intricate substructure in its age distribution within the subgroups themselves, reflected in the kinematic structure of the association: \cite{2018MNRAS.476..381W} concluded that there's no evidence to say that the three subgroups of Sco-Cen are the result of a global expansion of three small, independent clusters; rather, they argue for the existence of substructures, that they could not resolve due to the small sample size; such subgroups should not be the outcome of single, individual bursts of star formation, but rather of several minor subgroups, born independently of each other.

All the elements of our analysis converge towards a single solution: namely, that star formation in USCO must have lasted more than 10 Myr. 
After starting producing stars almost simultaneously with LCC ($\sim 15$ Myr ago), as confirmed by the presence of stars coeval to the latter \citep{2019aa...623A.112D}, the formation continued: the existence of B-type stars with clear indication of youth like $\tau$~Sco \citep{2015MNRAS.448.2737R} or HD 142184 \citep{2011MNRAS.410..190T} argues for some star formation occurred between the first outburst and the age of $\rho$ Ophiuchi. Whether this star formation was continuous or rather experienced a late burst is a different question: in this regard, the existence of a kinematic bimodality similar, albeit less pronounced, to that of Taurus \citep{2017ApJ...838..150K}, is significant, as it is intriguingly consistent with the late-burst scenario put forward by \cite{2017ApJ...842..123F}.

The intricacy of star formation in USCO lies in the fact that it happens in small formation episodes (a few 10s-100s stars each), as already pointed out by \citep{2016MNRAS.461..794P}, and that the older events have at present time lost their kinematic imprint. This is seen also in the compact populations in Lupus within LCC, where the stars are younger than the diffuse population not only of the surrounding LCC, but also of USCO \citep{2019aa...623A.112D}. The most compact population of Upper Scorpius, $\rho$ Ophiuchi, has a population of protostars still embedded in the molecular clouds --hence invisible to Gaia-- only 0.3-1 Myr old \citep{1999ApJ...525..440L,2008hsf2.book..351W}, with external, less extinct regions spanning between 2 and 5 Myr, as we have shown. The interesting fact is that the group is under strong influence from the rest of USCO \citep{2020aa...638A..74L}, with positive feedback on the star formation rate in the L1688 cloud \citep{2006MNRAS.368.1833N}, whose stars have not dispersed yet \citep{2017aa...597A..90D}; moreover, a possible echo of its formation related to a shock wave by a supernova might still be seen as a radial velocity gradient of $\sim 1.0$ km s$^{-1}$ pc$^{-1}$ \citep{2016aa...588A.123R}. An even more significant observation is that the bubble around the eponymous star $\rho$ Ophiuchi is inflating with a velocity $\sim 1.3$ km s$^{-1}$, so that the time needed to inflate the bubble ($d \approx 1.36$ pc) is $\sim 1.2$ Myr \citep{2016aa...592A..88P}. The isochronal ages of the members of this bubble is about $\sim 5-10$ Myr, much higher than the kinematic age of the bubble. \cite{2016aa...592A..88P} speculate that the assumption of a constant expansion is not valid in the first phases, when the dense material surrounding the young stars actively acts to delay the expansion.

This conclusion is directly reflected into what we have found in our work: kinematic ages of the subgroups of USCO are always smaller than isochronal ages, even neglecting issues inherent to age determination of low-mass stars that would further raise the discrepancy. The idea that stars in USCO formed well before their group started dispersing can be related to a initial bound state of the gas-rich structures, disrupted after the dispersal of gas by stellar feedback \citep{2014PhR...539...49K}. Therefore, the timescale of gas removal --quantified as 2-7 Myr for some nearby galaxies \citep{2021MNRAS.504..487K}-- could be intriguingly estimated as the difference between isochronal and kinematic ages.

\section{Conclusions}
\label{section:conclusions}

This work has shown that the star-forming region Upper Scorpius can be divided into two populations, carrying a distinct kinematic imprint. The existence of a clustered population cannot be attributable to a random concentration of sources, as shown with both a 2D and a 3D kinematic analysis. While the diffuse population does not appear to have been more concentrated in the past, the clustered population shows a clear tendency towards a more compact past configuration. We have further divided the clustered population in 8 groups, the most evident clustering at $\sim 4$ Myr ago.

This duality within USCO is clearly reflected in the age determinations obtained through isochrone fitting: the diffuse population is characterized by a flatter age distribution than the clustered population, whose relative youth is consistent with a late burst of star formation. The fraction of stars bearing mark of primordial discs is significantly higher for the latter ($f_D = 0.24\pm0.02$) than for the former ($f_D = 0.10\pm 0.01$).

Even if the absolute ages $t_P$ and $t_D$ provided in this work should be taken with caution due to known uncertainties in theoretical models, the relative ages are significant, and argue for a strong difference between the populations and, on a lesser extent, among the groups themselves. The star formation history in Upper Scorpius appears to have been heavily substructured, with several small events comprising at most a few hundreds of stars, and spread over $\sim 10$ Myr. The kinematic structure of the association is still visible inside the youngest part of the association, but has already been erased in the oldest. The systematic differences between kinematic and isochronal ages are likely due to the timescale of gas dispersal, intriguingly building a bridge from the early phases of star formation in molecular clouds to the final stages of star dispersal into the galactic field.

\section*{Acknowledgements}

We would like to thank the anonymous referee for the useful comments that helped us to refine the manuscript.

This work has employed data from the European Space Agency (ESA) space mission Gaia. Gaia data are being processed by the Gaia Data Processing and Analysis Consortium (DPAC). Funding for the DPAC is provided by national institutions, in particular the institutions participating in the Gaia MultiLateral Agreement (MLA). The Gaia mission website is https://www.cosmos.esa.int/gaia. The Gaia archive website is https://archives.esac.esa.int/gaia. Funding for the Sloan Digital Sky Survey IV has been provided by the Alfred P. Sloan Foundation, the U.S. Department of Energy Office of Science, and the Participating  Institutions. SDSS-IV acknowledges support and resources from the Center for High Performance Computing  at the University of Utah. The SDSS website is www.sdss.org. SDSS-IV is managed by the Astrophysical Research Consortium for the Participating Institutions of the SDSS Collaboration including the Brazilian Participation Group, the Carnegie Institution for Science, Carnegie Mellon University, Center for Astrophysics | Harvard \& Smithsonian, the Chilean Participation Group, the French Participation Group, Instituto de Astrof\'isica de Canarias, The Johns Hopkins University, Kavli Institute for the Physics and Mathematics of the Universe (IPMU) / University of Tokyo, the Korean Participation Group, Lawrence Berkeley National Laboratory, Leibniz Institut f\"ur Astrophysik Potsdam (AIP),  Max-Planck-Institut f\"ur Astronomie (MPIA Heidelberg), Max-Planck-Institut f\"ur Astrophysik (MPA Garching), Max-Planck-Institut f\"ur Extraterrestrische Physik (MPE), National Astronomical Observatories of China, New Mexico State University, New York University, University of Notre Dame, Observat\'ario Nacional / MCTI, The Ohio State University, Pennsylvania State University, Shanghai Astronomical Observatory, United Kingdom Participation Group, Universidad Nacional Aut\'onoma de M\'exico, University of Arizona, University of Colorado Boulder, University of Oxford, University of Portsmouth, University of Utah, University of Virginia, University of Washington, University of  Wisconsin, Vanderbilt University,  and Yale University. This work made use of the Third Data Release of the GALAH Survey (Buder et al 2021). The GALAH Survey is based on data acquired through the Australian Astronomical Observatory, under programs: A/2013B/13 (The GALAH pilot survey); A/2014A/25, A/2015A/19, A2017A/18 (The GALAH survey phase 1); A2018A/18 (Open clusters with HERMES); A2019A/1 (Hierarchical star formation in Ori OB1); A2019A/15 (The GALAH survey phase 2); A/2015B/19, A/2016A/22, A/2016B/10, A/2017B/16, A/2018B/15 (The HERMES-TESS program); and A/2015A/3, A/2015B/1, A/2015B/19, A/2016A/22, A/2016B/12, A/2017A/14 (The HERMES K2-follow-up program). We acknowledge the traditional owners of the land on which the AAT stands, the Gamilaraay people, and pay our respects to elders past and present. This paper includes data that has been provided by AAO Data Central (datacentral.aao.gov.au). This publication makes use of data products from the Two Micron All Sky Survey, which is a joint project of the University of Massachusetts and the Infrared Processing and Analysis Center/California Institute of Technology, funded by the National Aeronautics and Space Administration and the National Science Foundation. This publication makes use of data products from the Wide-field Infrared Survey Explorer, which is a joint project of the University of California, Los Angeles, and the Jet Propulsion Laboratory/California Institute of Technology, and NEOWISE, which is a project of the Jet Propulsion Laboratory/California Institute of Technology. WISE and NEOWISE are funded by the National Aeronautics and Space Administration. This research has made use of the SIMBAD database and Vizier services, operated at CDS, Strasbourg, France and of the Washington Double Star Catalog maintained at the U.S. Naval Observatory. Finally, we would like to thank NASA’s Astrophysics Data System. R.G. and D.M. acknowledge funding by the PRIN-INAF 2019 "Planetary systems at young ages (PLATEA) and the ASI-INAF agreement n.2018-16-HH.0.
M.B. acknowledges funding by the UK Science and Technology Facilities Council (STFC) grant no. ST/M001229/1.
\section*{Data Availability}

\textsc{madys} is soon to be made publicly available for use.

The data underlying this article can be freely accessible on Vizier at http://vizier.u-strasbg.fr/viz-bin/VizieR. Queries on the Gaia archive can be done using the dedicated web page: https://gea.esac.esa.int/archive/ for Gaia, .

The trace-back animations of our samples, described throughout this work, are provided as supplementary material to the paper.



\bibliographystyle{mnras}
\bibliography{biblio} 




\appendix

\section{Gaia DR2 corrected bp\_rp\_excess\_factor}
\label{section:excess_factor}

A known problem of Gaia's $G_{BP}$ and $G_RP$ photometry is the insensitivity to variation in local background levels, affecting the derived photometry especially for faint magnitudes \citep{2018aa...616A...3R}. For this reason, \citet{2018aa...616A...4E} put forward the idea of a quality metric, named {\it{BP-RP excess factor}}, which is the ratio of the sum of the two fluxes and the $G$–band flux:
\begin{equation}
C = \frac{F_{BP} + F_{RP}}{F_G}
\end{equation}
and derives its effectiveness from the behaviour of instrumental passbands and response; $C \approx 1$ for well-behaved sources. \citet{2020arXiv201201916R} show that the actual distribution of $C$ is more complex and colour-dependent, with larger expected values at redder colours, and introduce a {\it corrected BP-RP flux excess factor} $C^*$ defined as:
\begin{equation}
    C^* = C - f(G_{BP} - G_{RP})
\end{equation}
where $f(G_{BP} - G_{RP})$ is an appropriate piecewise polynomial function. $C^*$ is defined in such a way that its expected value is zero for well-behaved sources, and its value can be used as a way to discriminate between sources with {\it{good}} and {\it{bad}} $(G_{BP}, G_{RP})$ photometry. As the standard deviation of $C^*$ increases at fainter magnitudes, with standard deviation given by:
\begin{equation}
    \sigma_{C^*}(G) = c_0 C + c_1 G^m  
\end{equation}
they suggest to remove the sources such that $|C^*|>N \sigma(G)$.

\begin{figure}
\centering
\includegraphics[width=9cm]{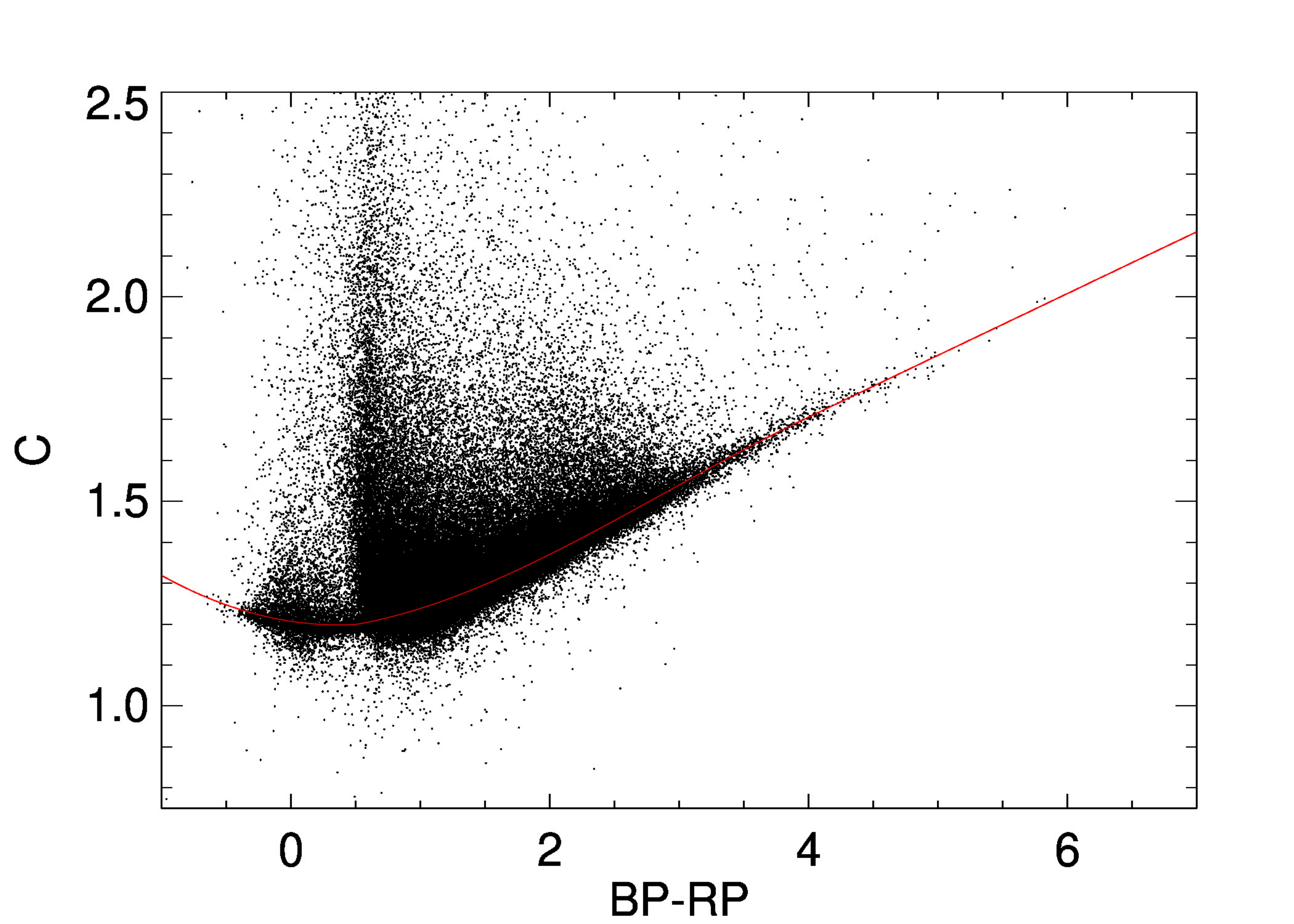}
\caption{Dependence of the bp\_rp\_excess\_factor on $G_{BP}-G_{RP}$ colour, using the set of standard stars by \protect\cite{2000PASP..112..925S}. The best-fitting polynomial is overplotted in red.}
\label{fig:exc_factor1}
\end{figure}

As a thorough analysis of this kind is not known to the authors for Gaia DR2 photometry, it was chosen to follow the same line of reasoning of \citet{2020arXiv201201916R}, but applied to the photometry of Gaia DR2. To start with, we recovered the set of standard stars by \cite{2000PASP..112..925S} that was used by \citet{2020arXiv201201916R}, comprising $\sim 200000$ stars. After reproducing their results for Gaia EDR3 photometry, we repeated the analysis for Gaia DR2. Reasoning in the same way as in their discussion, a piecewise polynomial was fitted to the data (Fig.~\ref{fig:exc_factor1}). However, an additional dependence on magnitude was observed in the corrected excess, showing up as a small counterclockwise rotation in the $(G,C^*)$ plane, well fitted by a straight line.

\begin{figure}
\centering
\includegraphics[width=9cm]{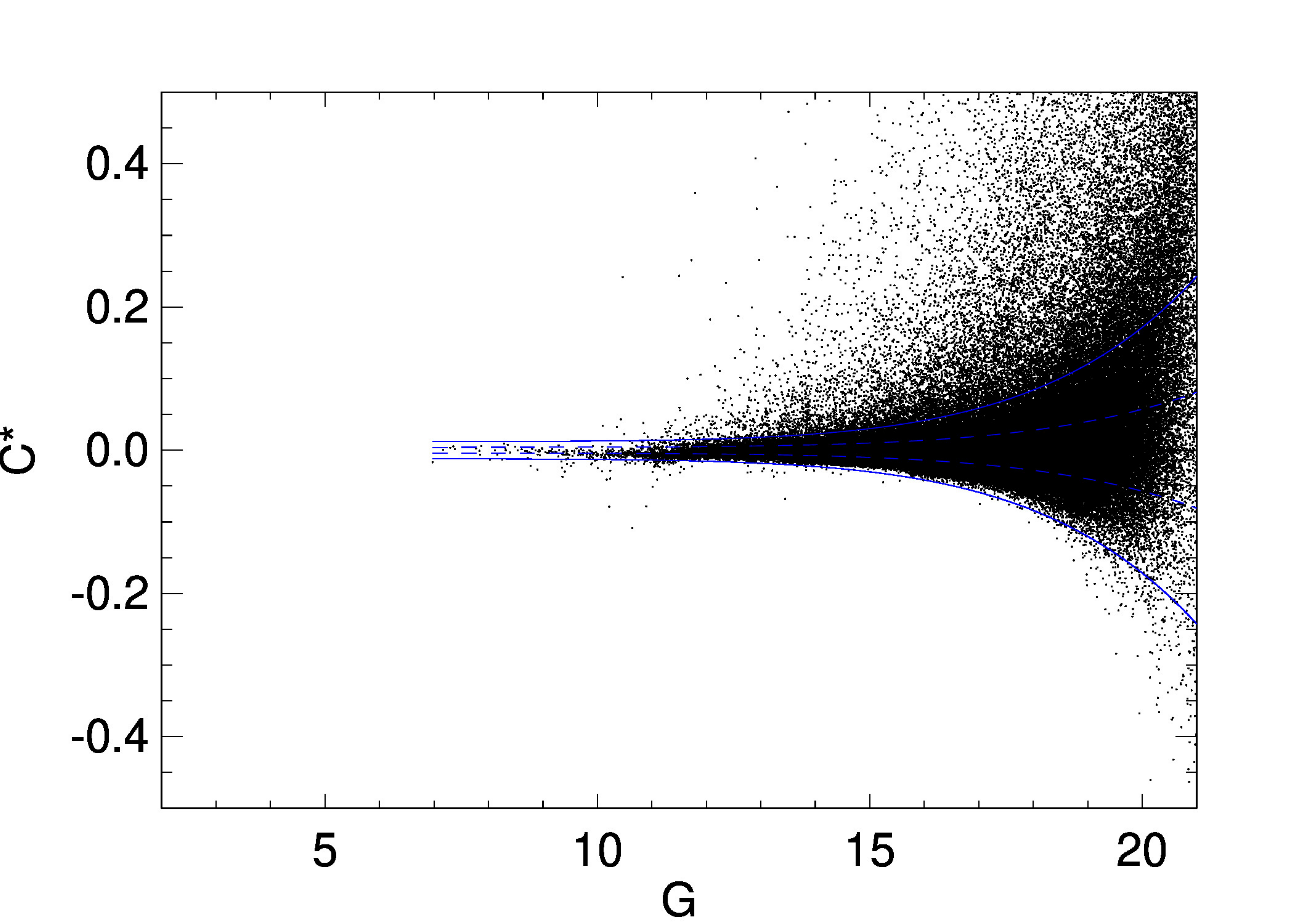}
\caption{Trend of $C^*$ with $G$ for the set of standard stars. As expected, the distribution widens out at fainter magnitudes. In blue, the $3 \sigma$ threshold that we applied to exclude sources with unreliable $(G_{BP},G_{RP})$ photometry.}
\label{fig:exc_factor2}
\end{figure}

The final equation defining the {\it{corrected BP-RP excess factor}} is given by:
\begin{equation}
\label{eq:excess_f}
C^* = C + a_0+a_1\Delta G+a_2 \Delta G^2+a_3 \Delta G^3+a_4 G
\end{equation}
where $\Delta G=(G_{BP}-G_{RP})$; numerical values for the constants are provided in Table ~\ref{tab:excess_param}. The distribution of $C^*$ peaks at about 0 for well-behaved sources at all magnitudes but, when considering subsamples of stars with similar brightness, it tends to widen out for fainter {\it{G}} (Fig.~\ref{fig:exc_factor2}); a varying standard deviation $\sigma_{C^*}(G)$ can be defined as:
\begin{equation}
\sigma_{C^*}(G)=0.004+8\times10^{-12} \times G^{7.55}
\end{equation}

\begin{table}
  \centering
   \caption{Best-fitting parameters for Eq.~\ref{eq:excess_f}}
   \begin{tabular}{c|ccc} \hline
& $\Delta G < 0.5$ & $ \text{ } 0.5 \leq \Delta G < 3.5$ & $\Delta G \geq 3.5$ \\
            \hline
$a_0$ & -1.121221 & -1.1244509 & -0.9288966 \\
$a_1$ & +0.0505276 & +0.0288725 & -0.168552 \\
$a_2$ & -0.120531 & -0.0682774 & 0 \\
$a_3$ & 0 & 0.00795258 & 0 \\
$a_4$ & -0.00555279&-0.00555279 & -0.00555279 \\
            \hline
   \end{tabular}
   \label{tab:excess_param}
\end{table}
\noindent

\begin{figure}
\centering
\includegraphics[width=9cm]{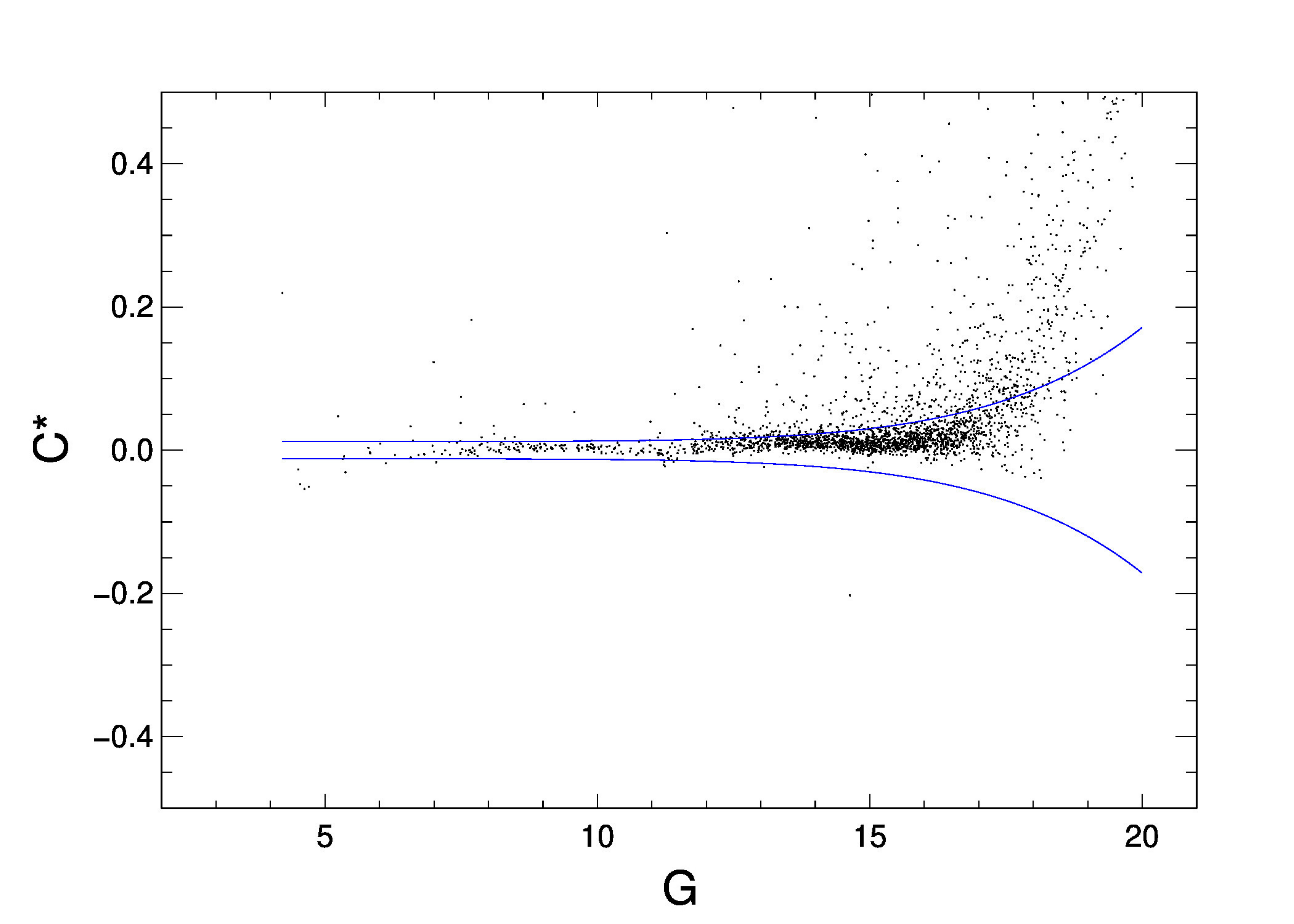}
\caption{Same as Fig.~\ref{fig:exc_factor2}, but using our 2D sample.}
\label{fig:exc_factor3}
\end{figure}

We decided to use a cut at $3 \sigma$, as it effectively excludes stars that look visibly far from their expected positions in the $(G_{BP}-G_{RP},G)$ diagram. The trend of $C^*$ with $G$ for our 2D sample is shown in Fig.~\ref{fig:exc_factor3}: 889/2745 sources ($32\%$) were flagged for unreliable ($G_{BP}$, $G_{RP}$) photometry. Given that most of these stars behave well in the $(G,J)$, $(G,H)$ colours, this filter yields a significant improvement of the quality of isochronal age estimates.

\section{Corrected proper motions}
\label{section:proj_effects}
Any non-zero motion of an association with respect to the Sun will manifest itself as a bias in the kinematic reconstruction.

Let us suppose, for instance, that a group of stars is rigidly approaching the Sun with velocity $v_{r,c}$. If we conveniently define a Cartesian frame $(\hat{x},\hat{y},\hat{z})$ with one axis, let's say $\hat{y}$, connecting the centre of the group to the Sun, we might say that all the stars have velocity $(0,v_{r,c},0)$. However, when we describe the position of individual stars on the sky plane, $v_{r,c}$ will be split in $(v_\alpha,v_\delta,v_r)$ depending on the relative $(\alpha,\delta)$ with respect to the centre of the association. It is easy to verify that this results in a contribution to proper motions, that are seen to 'escape' from the centre. This is referred to as {\it{virtual expansion}}, and reflects the simple idea that the approaching association will span, as time passes, a greater extension on the sky. In a similar way, any tangential motion of the group along $x$ and $z$ will contribute differently to $(v_\alpha,v_\delta,v_r)$ according to $(\alpha,\delta)$, producing a spurious velocity difference among group members.

This argument explains why, if interested in the analysis of peculiar motions within an association, one must estimate and subtract from observed proper motions the contribution coming from the bulk motion.

In order to take this factor into account, we began by looking for the centre of Upper Scorpius; starting from our 3D sample, we estimated the centre at approximately $(\alpha_c,\delta_c,r_c)=(244.55^\circ,-23.79^\circ,143.3$ pc$)$.

Then, we performed a coordinate transformation analogue to that of Eq.~\ref{eq:coord_transf1}:
\begin{equation} \label{eq:coord_transf2}
    \begin{cases}
    x=r \cos (\delta-\delta_c) \sin (\alpha-\alpha_c) \\
    y=r \cos (\delta-\delta_c) \cos (\alpha-\alpha_c) \\
    z=r \sin (\delta-\delta_c)
\end{cases}
\end{equation}
The relation among velocities is found by deriving:
\begin{equation}
\begin{pmatrix}
v_x \\
v_y \\
v_z 
\end{pmatrix} =
\begin{pmatrix}
 r \cos \Delta \alpha \cos \Delta \delta & -r \sin \Delta \delta \sin \Delta \alpha & \cos \Delta \delta \sin \Delta \alpha \\
-r \sin \Delta \alpha \cos \Delta \delta & -r \sin \Delta \delta \cos \Delta \alpha & \cos \Delta \delta \cos \Delta \alpha \\
0 & r \cos \Delta \delta & \sin \Delta \delta
\end{pmatrix}
\begin{pmatrix}
\mu_\alpha \\
\mu_\delta \\
v_r
\end{pmatrix}
\end{equation}
where we define $\Delta \alpha:=\alpha-\alpha_c$ and $\Delta \delta:=\delta-\delta_c$. We might write, in a compact form, $v_{cart}=A v_{eq}$. 
Let us consider the simplest case in which the peculiar motions are null, and the associations moves rigidly in the sky with constant $(V_x, V_y, V_z)$. The reflection of the bulk motion in equatorial coordinates is given by $v_{eq}=A^{-1} v_{cart}$, i.e.:
\begin{equation} \label{eq:bulk_corr}
\begin{pmatrix}
\mu_\alpha \\
\mu_\delta \\
v_r
\end{pmatrix} =
\begin{pmatrix}
 \frac{\cos \Delta \alpha}{r \cos \Delta \delta} & \frac{\sin \Delta \alpha}{-r \cos \Delta \delta} & 0 \\
\frac{\sin\Delta \alpha \sin \Delta \delta}{-r} & \frac{\cos \Delta \alpha \sin \Delta \delta}{-r} & \frac{\cos \Delta \delta}{r} \\
\sin \Delta \alpha \cos \Delta \delta & \cos \Delta \alpha \cos \Delta \delta & \sin \Delta \delta
\end{pmatrix}
\begin{pmatrix}
v_x \\
v_y \\
v_z 
\end{pmatrix}
\end{equation}
In order to get consistent $\mu_\alpha^*=\mu_\alpha \cdot \text{cos} \delta$, $\mu_\delta$ [mas yr$^{-1}$] and $v_r$ [km s$^{-1}$], a conversion factor $p=1000/4.74$ must be used in the equations for proper motion components.

We estimate bulk Cartesian velocities as the median velocity components of the 3D sample: $(V_X, V_Y, V_Z)=(-7.20 \pm 0.01,-4.58 \pm 0.02,-16.29 \pm 0.01)$ km s$^{-1}$, where the errors are computed with $N=10000$ realisations of the same Monte Carlo approach employed throughout this work. The velocity dispersion, computed using the 16th and 84th percentiles of the distribution, corresponds to $(\sigma_{V_X},\sigma_{V_Y},\sigma_{V_Y})=([-1.98,+2.62],[-1.64,+1.61],[-1.25,+0.50])$ km s$^{-1}$, for a total 3D dispersion $\sim 3$ km s$^{-1}$.

To directly compare our results to those obtained by previous studies, we converted our velocities into the standard galactic right-handed frame UVW, where the origin lies in the Sun, $X$ heads towards the Galactic centre, $Y$ follows the Galactic rotation and $Z$ is directed towards the Galactic North Pole. We find $(U, V, W) = (-4.788 \pm 0.019, -16.378 \pm 0.015, -6.849 \pm 0.016)$ km s$^{-1}$, comparable with both \cite{2018MNRAS.477L..50G} and \cite{2012ApJ...758...31L}.

Subtracting the first two equations of Eq.~\ref{eq:coord_transf1} from proper motions from Gaia EDR3, we have shifted to a reference frame jointed to USCO, so that only peculiar motions are left. Projection effects of peculiar motions are not eliminated (see footnote ~\ref{foot:cartesian}), but it's the best we can achieve without possessing radial velocities for all the stars.


\bsp	
\label{lastpage}
\end{document}